\documentclass[fleqn,usenatbib,useAMS]{mnras}

\usepackage{graphicx}	
\usepackage{amsmath}	
\usepackage{multicol}        
\usepackage{subcaption}
\usepackage{bm}		
\usepackage{pdflscape}	
\usepackage{wasysym}
\usepackage{enumitem}

\usepackage[T1]{fontenc}
\usepackage{ae,aecompl}

\usepackage{newtxtext,newtxmath}
\newcommand{\orcid}[1]{\href{https://orcid.org/#1}{\includegraphics[width=8pt]{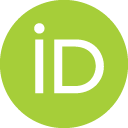}}}

\title[Fixed points II]{Irregular Fixation II: The orbits of irregular satellites}

\author[E. Grishin]{Evgeni Grishin$^{1,2}$ \orcid{0000-0001-7113-723X}%
\thanks{Contact e-mail: \href{evgeni.grishin@monash.edu}{evgeni.grishin@monash.edu}}%
\\
$^{1}$School of Physics and Astronomy, Monash University, Clayton, VIC 3800, Australia\\
$^{2}$OzGrav: Australian Research Council Centre of Excellence for Gravitational Wave Discovery, Clayton, VIC 3800, Australia}

\pubyear{2024}

\begin{document}
\defcitealias{gri24I}{paper I}

\label{firstpage}
\pagerange{\pageref{firstpage}--\pageref{lastpage}}
\maketitle

\begin{abstract} Irregular satellites (IS) are believed to have been captured during the Solar system's dynamical history and provide clues for the Solar system's formation and evolution. IS occupy a large fraction of the Hill sphere of their host planet and their orbits are highly perturbed by the Sun. We use a novel formalism developed in \citetalias{gri24I} to characterise their orbits in terms of an effective secular Hamiltonian (the Brown Hamiltonian) that accounts for their large orbital separations. We find that prograde satellites generally follow the Brown Hamiltonian, while retrograde satellites (which extend further) deviate more significantly. Nevertheless, the phase portrait is much better described by the Brown Hamiltonian for all satellites. We construct a semi-analytic criterion that predicts the librating orbit based on the effective energy due to the Brown Hamiltonian. We also check our results with highly accurate N-body integrations of satellite orbits, where initial conditions are loaded directly from the updated ephemeris from the NASA Horizons database. Although the retrograde librating orbits occupy more area in the parameter space, the vast majority of librating IS are prograde. Using our method we find $13$ librating satellites, $8$ of them previously known to librate, and the rest shown to librate for the first time. Further observations of existing and new satellites could shed more light on the dynamical history of the Solar system and satellite formation and test our results.
\end{abstract}





\section{Introduction} \label{sec:intro}

Irregular satellites (IS) are small objects which are weakly bound to their planet hosts. The main source of perturbations of their orbits is the Sun and other planets, rather than the inner quadrupolar distortion of the planets' oblateness and inner satellite system \citep{nesvorny03}. IS orbits usually have large eccentricities and inclinations, which vary significantly over time. Studying the orbits of IS is vital for understanding the origin and dynamical evolution of the Giant planets and the Solar system as a whole \citep{car02}.

 The vast majority of IS orbits are retrograde (with the mutual inclination $\iota$ between the two orbital planes being $> 90$ deg), which also exceeds further than prograde orbits (where $\iota< 90$ deg). The extended stability limit for retrograde orbits is due to the Coriolis force which counteracts the tidal shear from the Sun \citep{i79,i80}. This dichotomy is even more striking due to the lack of near-polar orbits near $\iota \sim 90$ deg \citep[e.g.][]{car02, nesvorny03}. 

Due to their large orbits, the dynamics of IS have been studied in terms of the long-term evolution of three bodies \citep{hb91, car02}. \cite{cb04} developed an effective secular theory that accounts for the relatively large period ratio between the satellite and the planet's orbit. {\cite{frouard11} studied the chaotic evolution of IS in the context of secular and mean motion resonances between the different giant planets}. The critical stability limit for arbitrary mutual inclination has been recently studied \citep{grishin2017,tory22} and found that the polar orbits are the least stable. The early instability of polar orbits arises due to the von-Zeipel-Lidov-Kozai effect \citep[ZLK,][]{vZ1910,lid62,koz62}, where periodic oscillations coherently change the eccentricity and inclination of the inner orbit on longer, secular, timescales.

In addition to destabilising polar orbits, the ZLK effect is caused due to a fixed point in the eccentricity $e_1$ -- argument of periapse $\omega_1$ phase space, and an orbit can potentially librate around it with limited range for $\omega_1$, which is different from apsidal advance where $\omega_1$ completes full revolutions. With hundreds of IS found to date, only a handful are found to be librating: Carpo and Euporie around Jupiter \citep{brozivic17jup, sheppard23}, Kiviuq, Ijiraq, and $\rm S2004\_S31$ around Saturn \citep{jacobson22}, Margaret around Uranus \citep{brozovic09,brozovic22}, and Sao and Neso around Neptune \citep{brozovic11}.  Moreover, \cite{car02} identified hypothetical librating families at higher inclinations. 

IS form a sub-class of hierarchical triple systems, where the inner binary of semi-major axis $a_1$ is perturbed by a distant tertiary at semi-major axis $a_2 \gg a_1$. For a small period ratio $P_1/P_2 \ll 1$, an integrable, analytical, approximate solution for the ZLK effect can be found \citep{kinoshita07}. The latter approximation, however, is inaccurate for orbits with non-negligible period ratio, where the system is still stable, but the timescale hierarchy is mild \citep{luo16, gpf18, man22}.

Expansions of the disturbing function of systems of 'mild' hierarchy have a long history, from the Lunar theory \citep{brown1, brown3, brouwer61} to multiple stellar \citep{brown2, sod75} and compact object \citep{luo16, will21} systems.  \cite{tremaine23} had recently shown that the aforementioned (and other related) derivations are equivalent to the first instance of the 'Brown Hamiltonian', due to \cite{brown1}. {Generalised analogues
of Brown’s Hamiltonian for octupole and higher order potentials are
given by \cite{lei18}}.

{For even larger period ratios, even the corrected double-averaged theory is insufficient, and additional higher-order corrections are required. \cite{beauge06} used a canonical perturbation theory while \cite{lei19} used a literal expansion of the disturbing function. In both cases, the averaging procedure is done \textit{after} the canonical transformation of the osculating elements. This is contrary to the Brown Hamiltonian formalism, where first a single averaging over the inner orbit is applied before the canonical transformation \citep{luo16, tremaine23}, which gives more accurate results where the difference between the outer and inner period are blurry. While Brown's Hamiltonian is valid for any eccentricity, the higher-order expansion relies on Legendre expansions eccentricities in terms of Hansen coefficients, which does not converge for eccentricity above $e_c=0.66$ (see end of sec. 2 of \citealp{beauge06}, and also \citealp{wintner41, lei19}).}

In the first paper \citep[][hereafter \citetalias{gri24I}]{gri24I} we derived an analytical expression for fixed points for the Brown Hamiltonian, which depend on dimensionless quantities constructed from the triple system's architecture. We also explored the available parameter space for librating orbits if they are initialised at the fixed point. The energy associated with the Brown Hamiltonian is conserved on average, while the fluctuating terms are also proportional to the period ratio.  

In this second paper in the series, we apply our results to explore the orbits of IS. We formulate an effective energy condition for the orbits to be allowed to librate. We also confirm that the known librating IS satisfy this condition and suggest that several other satellites are librating. 

The paper is organised as follows: We first briefly review the main results of \citetalias{gri24I} in sec. \ref{s2}. We describe how to select IS satellites from the NASA JPL data in sec. \ref{s3}. In sec. \ref{s4} we apply our result to IS of the giant planets. We first devise a condition for orbits to be potentially librating in \ref{s41}, and refine it in \ref{s2}. The condition is based on the conservation of the effective energy due to the Brown Hamiltonian and on the amplitude of the osculating fluctuations of the orbital elements. In sec. \ref{s5} we test it with direct N-body integrations and compare the orbital evolution of various ISs. Using this condition we find all $8$ known librating satellites and suggest that $5$ others are also librating. Finally in sec. \ref{s6} we conclude our results and discuss our limitations and potential avenues for future research directions.

\section{Fixed points and librating zones} \label{s2}

 Here we briefly summarise the main results of \citetalias{gri24I}.

Consider a hierarchical system with masses $m_0,m_1 \ll m_2$, orbital separations $a_1 \ll a_2$ and other orbital elements denoted by subscript $1$ for the inner orbit and $2$ for the outer orbit. In the quadrupole approximation, the quantity $j_z \equiv \sqrt{1-e_1^2}\cos\iota$ is conserved. The motion of the satellite can be described by two terms. The first is the secular quadrupole Hamiltonian
 \begin{align}
{\mathcal{H}}_{{\rm sec}} & \equiv\mathcal{\langle\langle H}\rangle\rangle=C\left[1-6e_{1}^{2}-3j_z^2 +15e_{1}^{2}\sin^{2} \iota \sin^{2}\omega_{1}\right]\label{eq:sec_quad}\nonumber \\
C & \equiv\frac{G\mu_{\rm in} m_2}{8a_{2} \left(1-e_{2}^{2}\right)^{3/2}}\left(\frac{a_{1}}{a_{2}}\right)^{2},
\end{align}
where $\mu_{\rm in}=m_0 m_1/(m_0+m_1)$ is the reduced mass of the inner binary, $\omega_1$ is the argument of pericentre and $\iota$ is the mutual inclination between the orbits.

The second term is the Brown Hamiltonian
 \begin{equation}
     {\mathcal{H}}_{\rm B}=-\epsilon_{{\rm SA}}C\frac{27}{8}j_{z}\left(\frac{(1-j_{z}^{2})}{3}+8e_1^{2}-5e_1^2 \sin^2\omega_1 \sin^2\iota \right)  + \mathcal{O}(e_2^2), \label{h_Brown} 
 \end{equation}
Here,
\begin{equation}
    \epsilon_{\rm SA} = \left(\frac{a_1}{\ell_2}\right)^{3/2} \left( \frac{m_2^2}{(m_0+m_1)(m_0+m_1+m_2)} \right ) ^{1/2}\label{eps_sa}
\end{equation}
 is the parameter that measures the relative accuracy of the double averaging approximation which is roughly equal to the period ratio $P_1/P_2$, $\ell_2 = a_2(1-e_2^2)$ is the semi-latus rectum. The total Hamiltonian is $\mathcal{H}_{\rm tot} = {\mathcal{H}}_{{\rm sec}} + {\mathcal{H}}_{\rm B} $. 

In \citetalias{gri24I} we found that the fixed points are given in terms of $j_z$ and $\epsilon_{\rm SA}$ by $\omega_{\rm fix} = \pi/2, 3\pi/2$ and \begin{align}
    e_{\rm fix}=\sqrt{ 1 - \sqrt{ \frac{5(1+\frac{9}{8}\epsilon_{\rm SA}j_z)}{3(1-\frac{9}{8}\epsilon_{\rm SA}j_z)} } |j_z|} \label{efix}
\end{align}
The associated inclination at the fixed point is
\begin{equation}
    \cos \iota_{\rm fix} = \frac{j_z}{x^{1/2}}= {\rm sign}(j_z) \left( \frac{3}{5} \right)^{1/4} \sqrt{\frac{1-\frac{9}{8}\epsilon_{\rm SA} j_z}{{1+\frac{9}{8}\epsilon_{\rm SA} j_z}} }|j_z|^{1/2}. \label{ifix}
\end{equation}
Note that for $\epsilon_{\rm SA}=0$ we retain the fixed point previously found for the secular ZLK solution \citep[only for $\mathcal{H}_{\rm sec}$, ][]{antognini15,hamers-zlk}.

A direct N-body integration shows that the energy given by $\mathcal{H}_{\rm tot}$ is conserved on average, and fluctuates about a mean value with an amplitude proportional to $\epsilon_{\rm SA}$. Using a large grid of initial conditions, we initialised $\sim 3\times 10^5$ three body systems at the fixed points according to Eq. \ref{efix} (if they exist) in \citetalias{gri24I}. We then integrated the systems and classified whether they were able to stay in libration. Besides the fractal boundary near the instability, we found a numerical fit for the boundary between libration and circulation in $j_z$ -- $\epsilon_{\rm SA}$ space:
 \begin{equation}
    \epsilon_{{\rm SA}}(j_{z})=\begin{cases}
0.4(\sqrt{3/5}-j_{z})^{0.7} & j_{z}>0;\ \epsilon_{{\rm SA}}\in[0.01,0.16]\\
0.08+0.5(0.8+j_{z})^{0.52} & j_{z}<0;\ \epsilon_{{\rm SA}}\in[0.08,0.23]\\
0.08-0.54(0.8+j_{z})^{0.52} & j_{z}<0;\ \epsilon_{{\rm SA}}\in[0.01,0.08]
\end{cases}, \label{eps_fit}
\end{equation}
Note that the retrograde case is multivalued.

In the remainder of the paper, we describe the irregular satellite database and the location of the known librating satellites in the   $j_z$ -- $\epsilon_{\rm SA}$ space. We then describe our semi-analytic criterion for potentially librating orbits and show how the orbital evolution can be quantitatively described by the Brown Hamiltonian. 

\section{Initial satellite selection}\label{s3}

Currently, the NASA Horizons database has $293$ satellites. Although the \href{https://ssd.jpl.nasa.gov/sats/elem/#refs}{Planetary Satellite Mean Elements database} has listed orbital elements, they are based on precessing ellipse models, but could significantly differ from the osculating orbital elements. The mean elements are descriptive and can not be used as initial conditions for N-body integrations (Marina Brozivi{\'c}, private communication).

In order to load the initial conditions for the satellites, we use a \href{https://rebound.readthedocs.io/en/latest/ipython_examples/Horizons/}{built-in pipeline} which directly reads and initialises an N-body system with the most up-to-date state vector provided by Horizons. The pipeline is publicly available as part of the \href{https://rebound.readthedocs.io}{\texttt{Rebound}} code, which is a highly accurate N-body code \citep{rein12}. All the integrations in sec. \ref{s5}  were computed using the \texttt{IAS15} adaptive timestep integrator, accurate to machine precision for billions of orbits \citep{ias15} and is built-in within \texttt{Rebound}.

After setting up three body systems, we calculate the orbits and use the calculated (osculating) orbital elements. We filter only as satellites of the giant planets, which excludes the Moon, Mars' two satellites and Pluto's five satellites. We then take only IS as follows: For each giant planet we first calculate its Laplace radius \citep{tremaine09}
\begin{equation}
  r_L = \left(J_2' R_p^2 a_p^3 (1-e_p^2)^{3/2} \frac{m_p}{M_\odot} \right)^{1/5},  \label{laplace}
\end{equation}
where $J_2'$ is the quadrupolar distortion coefficient, $R_p, a_p, e_p$ and $m_p$ are the giant planets' radius, semi-major axis, eccentricity and mass, respectively.  We will use the values for $J_2'$ which includes the contribution from the inner satellite system, found in table 1 of \cite{tremaine09}. Satellites with orbits $r<r_L$ will be subjected to the inner quadrupole, which causes apsidal advance and prevents ZLK oscillations \citep{lml15, gri18b, gri20}.

We deem the satellite irregular if $a_1 > r_L$, which leaves us with $228$ objects. This ensures that the dynamics are governed by the outer quadrupolar perturbations from the Sun, and not the inner quadrupolar distortion.

\section{Energy conditions}\label{s4}

\subsection{Instantaneous energy condition}\label{s41}
For each satellite, we calculate $j_z$ and $\epsilon_{\rm SA}$ (from the osculating elements), and also the current energy:

\begin{equation}
    E_c = \mathcal{H}_{\rm sec}(j_z, e_1, \omega_1) + \mathcal{H}_B(\epsilon_{\rm SA}, j_z, e_1, \omega_1) \label{E_c}
\end{equation}
{We note that the osculating $j_z$ and $\epsilon_{\rm SA}$ are different from their averaged values, $\bar{j}_z$, $\bar{\epsilon}_{\rm SA}$. Although $j_z$, $\epsilon_{\rm SA}$ are oscillating, we will treat them as constant for now. This first step is done to filter out orbits which can not be librating, from 'suspected' orbits.}

The total Hamiltonian essentially has one degree of freedom for the conjugate variables $e_1$\footnote{More precisely, the conjugate Delaunay variable is $G_1 =\mu_{\rm in} \sqrt{Gm_p a_1 (1-e_1^2)}$, which corresponds to the orbital angular momentum, see details in \citetalias{gri24I}.} and $\omega_1$. The mutual inclination $\iota$ is eliminated via $\cos \iota = j_z/ \sqrt{1-e_1^2}$.

We also calculate the critical energy for the separatrix -- the critical constant energy curve that separates circulating and librating orbits, which is evaluated at the origin:
\begin{equation}
    E_s = \mathcal{H}_{\rm sec}(j_z, 0,0) + \mathcal{H}_B(\epsilon_{\rm SA}, j_z, 0,0). \label{E_s}
\end{equation}
In practice, $e_1=0$ is an unstable hyperbolic fixed point, so we take $e_1=10^{-3}$ and $\omega_1=0$ to estimate $E_s$ numerically. An orbit is suspected as librating if $E_c > E_s$. Applying this procedure results in $33$ satellites suspected as librating. {It is important to stress that this is not a sufficient condition, since the osculating elements vary over time and the energy contours generated from the osculating elements are inaccurate.}
\begin{figure*}
    \centering

   \includegraphics[width=0.5\textwidth]{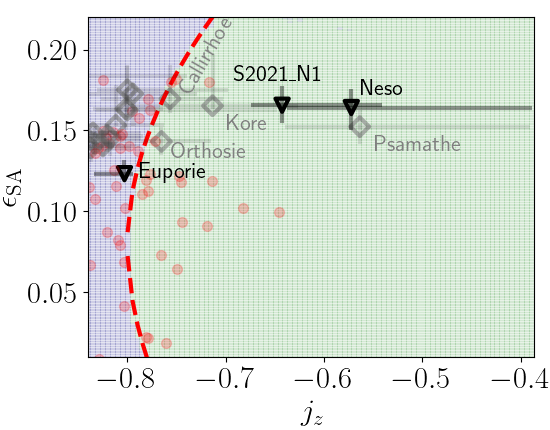}
   \includegraphics[width=0.485\textwidth]{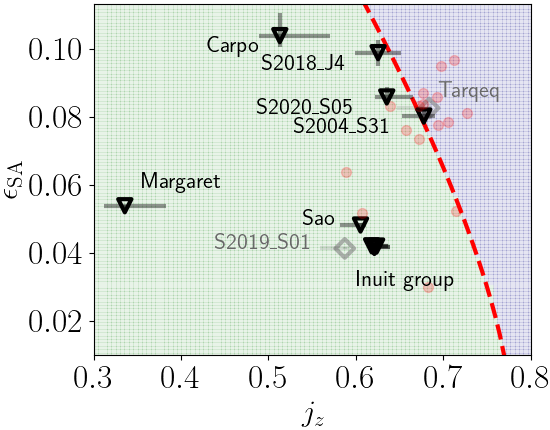}
   
    \caption{ IS in the $\epsilon_{\rm SA}$ - $j_z$ space. The background colours represent the regions of possible librating zones (green) for fixed points, and for regions where orbits which were initialised at the fixed point were circulating (blue), see Fig. 6 in \citetalias{gri24I}. The dashed red lines show the numerical fit to the boundary between circulation and libration (Eq. \ref{eps_fit}).  Red circles are IS which have $E_c<E_s$, so they are unlikely to librate. Grey diamonds are satellites that have $E_c>E_s$ but are found to be circulating from N-body simulations. Black triangles represent the $13$ satellites that have $E_c>E_s$ and are found to be librating in N-body simulations. $4$ of them belong to the inner Inuit group which shares similar orbital elements. {The variations in $\epsilon_{\rm SA}$ and $j_z$ for suspected and librating satellites from N-body integrations are shown in black horizontal and vertical lines.}}
    \label{fig1}
\end{figure*}

We integrate all the IS' orbits to check whether or not the criterion predicts a ZLK librating state. For simplicity, we set up three body systems of the Sun, the planet and the test particle, which is justified if the IS are beyond the Laplace radius. We show in sec \ref{s5} that the known librating satellites have a similar orbital evolution to what had been reported in the past, thus we believe that our approximate integrations capture the essence of the librating satellites.

In Fig. \ref{fig1} we show the result of our analysis for the prograde case (right) and the retrograde case (left). We find that $13$ of the $33$ suspected satellites are librating. All prograde suspected satellites are in the allowed region for libration, while many of the retrograde suspected satellites are outside the allowed region. {Note that there's no restriction where satellites can circulate in $\epsilon_{\rm SA} - j_z$ space (it depends on $\omega_1$), but libration is allowed only inside the green area.} $10$ of the librating satellites are prograde while only $3$ are retrograde. On the other hand, only $2$ of the suspected prograde satellites are circulating, and the other $18$ suspected and circulating orbits are retrograde.  

{We stress that the red lines are an empirical boundary which is obtained in \citetalias{gri24I}, and does not come from the Brown Hamiltonian (although see the appendix in \citetalias{gri24I} for a comparison to boundaries obtained from the Brown Hamiltonian). We see that besides Euporie, all the librating satellites are within the boundary, and many of the prograde ones lie close to it. Librating orbits move on stable islands and exhibit lower eccentricity variations, hence they are expected to be more stable, which is consistent with our findings.}

The main reason is that the retrograde orbits extend to much further distances (and larger $\epsilon_{\rm SA}$), and additional perturbations cause the energy condition to be less accurate. 
{High order theory could in principle remedy the discrepancy. In particular, \cite{lei19}'s model incorporates high-order secular theory which has good correspondence. However, \cite{lei19}'s model is hard to realise in practice due to its inherent complexity. Moreover, the literal expression of the separation in terms of mean anomalies and Hansen coefficients does not converge for eccentricity above $e_c=0.6627$ \citep{wintner41}. While Neso's eccentricities frequently exceed this value, $\rm S2021\_N1$'s eccentricity is marginal and Euporie's eccentricity is relatively far from this threshold. It is possible in principle to construct a secular theory somewhere between the Brown Hamiltonian's simplicity and \cite{lei19}'s accuracy, but it is beyond the scope of this paper.}

{Inspecting the error bars in Fig. \ref{fig1} we observe that the fluctuation in $j_z$ is much larger than in $\epsilon_{\rm SA}$. A typical estimate of the fluctuations is $2\Delta j_z$, where $\Delta j_z$ is given by from Eq. \ref{djz}, and is around $\propto 15 \epsilon_{\rm SA} e_1^2/4$. The proportionality constant is between $0.7-0.9$ and depends on the remaining orbital elements. The largest variation of $j_z$ is associated with the most eccentric satellites such as Neso and $\rm S20201\_N1$. For prograde satellites, Margaret and Carpo have also similar variation in $j_z$, since the ratio square of the maximal (or average) eccentricities $(0.9/0.65)^2\sim 1.9$ is comparable to the ratio of $\epsilon_{\rm SA} \sim 1.95$.  }

The latter diversity requires an additional examination and refinement of our condition, which is explored below.

\subsection{Mean energy condition}\label{s42}
To better quantity the possible orbits, we construct the following analysis: For each satellite, we calculate the dimensionles energy
\begin{equation}
    \mathcal{E}_c = \frac{E_{\rm c} - E_{\rm s}}{E_{\rm fix} - E_{\rm s}}. \label{curlyE}
\end{equation}
Here,
\begin{equation}
    E_{\rm fix} = \mathcal{H}_{\rm sec}(j_z, e_{\rm fix}, \pi/2) + \mathcal{H}_B(\epsilon_{\rm SA}, j_z, e_{\rm fix}, \pi/2) \label{E_c}
\end{equation}
is the energy at the associated fixed point, and it is the maximal effective energy the satellite can have in its current orbit.

For orbits which are suspected as librating, $\mathcal{E}_c>0$. The normalisation by $E_{\rm fix} - E_{\rm s}$ also guarantees that $\mathcal{E}_c \le 1$, where equality is achieved only at the fixed point.

In order to account for potential fluctuations, we estimate the variation in the energy via the fluctuation in $j_z$. We denote

\begin{equation}
    E_c^{\pm} = \mathcal{H}_{\rm sec}\left(j_z \pm \frac{\Delta j_z}{2}, e_1, \omega_1\right) + \mathcal{H}_B\left(\epsilon_{\rm SA}, j_z \pm \frac{\Delta j_z}{2}, e_1, \omega_1\right), \label{Ec_pm}
\end{equation}
where  the fluctuation $\Delta j_z$ is given by Eq. 35 of \cite{luo16}
\begin{equation}
    \Delta j_z = \frac{3 \epsilon_{\rm SA}}{8}\left(5e_{1}^{2}\left(\cos^{2}\omega_1+\frac{j_{z}^{2}}{1-e_{1}^{2}}\sin^{2}\omega_1\right)+1-e_{1}^{2}-j_{z}^{2}\right). \label{djz}
\end{equation}
Note that in the limit of high eccentricity (and $\sin\omega_1=1$) we get back to Eq. 25 in \cite{gpf18}
\begin{equation}
    \lim_{e_{1}\to1,j_{z}\to0}\Delta j_{z}=\frac{15}{8}\epsilon_{{\rm SA}}e_{1}^{2}\cos^{2}\iota_{{\rm min}}=\frac{9}{8}\epsilon_{{\rm SA}}e_{1}^{2}.
\end{equation}

From $E_c^\pm$ we can construct the relative energy fluctuation
\begin{equation}
    \mathcal{E}^\pm_c = \frac{E_{\rm c}^\pm - E_{\rm s}}{E_{\rm fix} - E_{\rm s}} \label{curlyEd}.
\end{equation}
Without loss of generality, we rearrange $\mathcal{E}_c^\pm$ such that  $\mathcal{E}_c^+ > \mathcal{E}_c^-$.

\begin{figure*}
    \centering

   \includegraphics[width=1.0\textwidth]{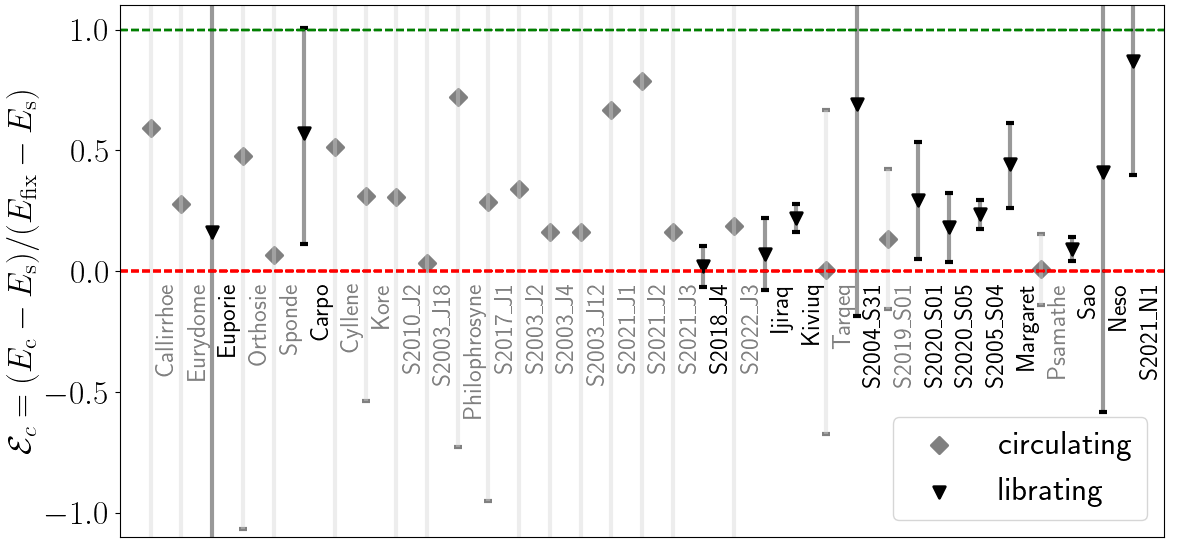}
  
    \caption{Dimensionless energy $\mathcal{E}_c$ of the suspected orbits. The marker styles and colours correspond to Fig. \ref{fig1}. The relative variations in the energies are given by $\mathcal{E}_c^\pm$. The dashed red line is the separatrix, while the dashed green line is the maximal energy at the fixed point, $\mathcal{E}_c=1$.}
    \label{fig2}
\end{figure*}

\begin{figure}
    \centering

   \includegraphics[width=0.5\textwidth]{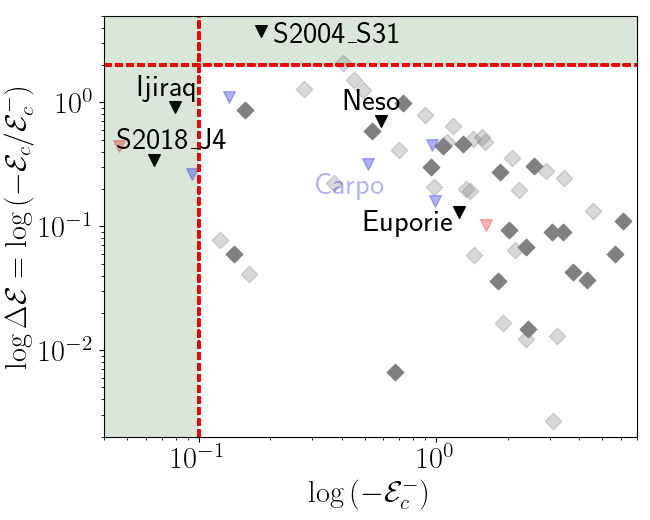}
   
    \caption{  The minimal energy versus the relative energy fluctuation. Only satellites with negative minimal energy (i.e. positive $-\mathcal{E}_c^-$ are shown). Dashed red lines indicate the boundaries between both orbit types at $-\mathcal{E}_c^- = 0.1$ and $\Delta \mathcal{E}_c = 2$. {The energy estimate for the librating orbits is done for several choices of $j_z$: $j_z$ from osculating elements (black), $j_z+\Delta j_z$ (blue) and $j_z - \Delta j_z$ (red). Semi-transparent grey diamonds are circulating satellites with $j_z \pm \Delta j_z$. The light green area is where only librating satellites are found}.}
    \label{fig3}
\end{figure}

\begin{figure*}
    \centering

    \includegraphics[width=1\textwidth]{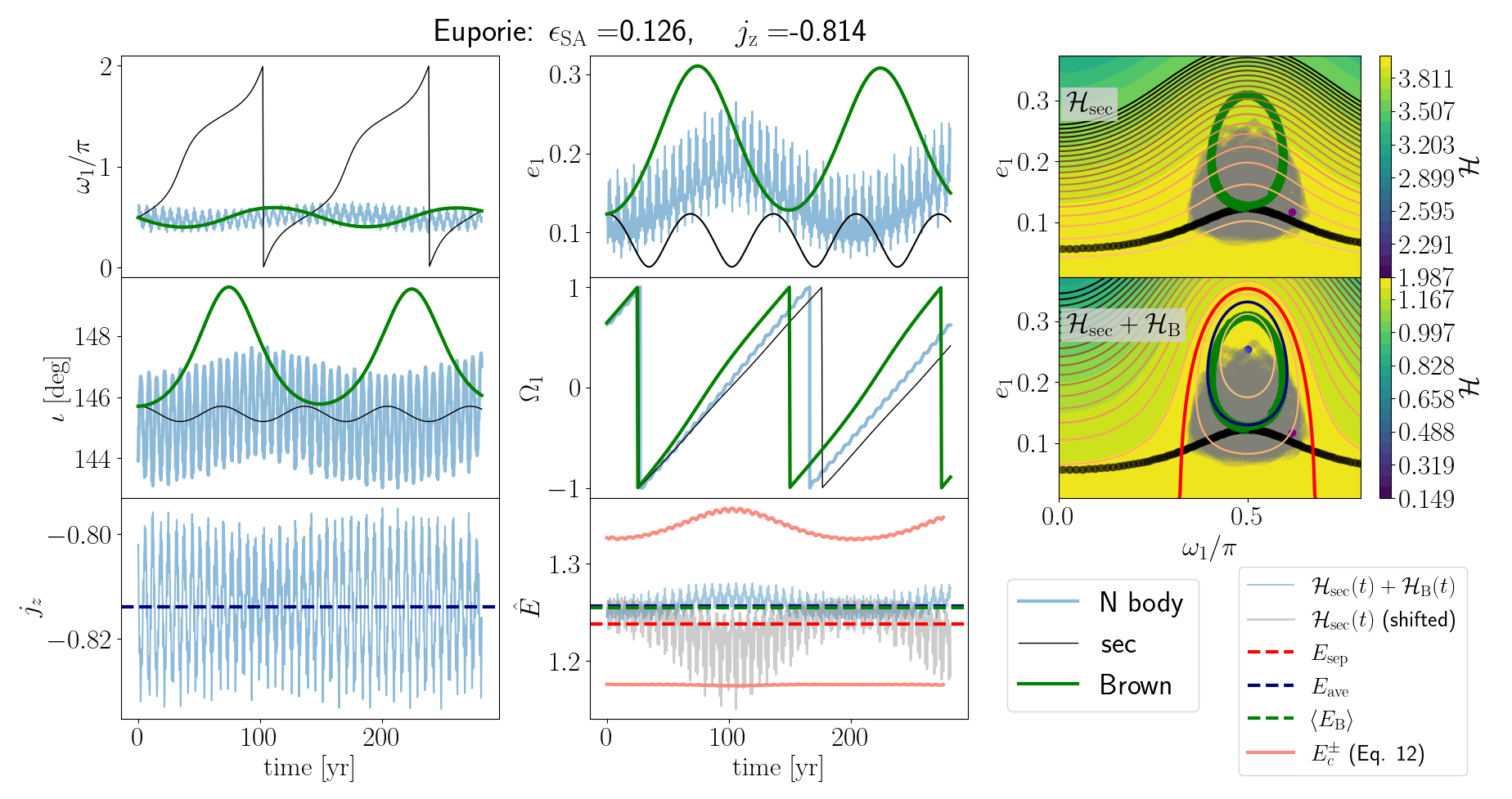}

     \caption{ Orbital elements of {Euporie.}  {Left} panels: Argument of pericentre $\omega_1$, Mutual inclination $\iota$, {and $j_z$. The light blue lines are from N-body integrations, the thin black line is from the secular evolution for $\mathcal{H}_{\rm sec}$, and the thick green line is for the secular evolution with the Brown Hamiltonian addition $\mathcal{H}_{\rm sec} + \mathcal{H}_{\rm B}$}. {Middle panels:} Eccentricity $e_1$, and angle of ascending node $\Omega_1$, normalised energy $\hat{E}=\mathcal{H}/C$. For the normalised energy, the dashed red line is the separatrix, and the dashed navy line is the average energy during an N-body integration, {and the dashed green line is the energy for the initial condition for the secular evolution code (averaged for one inner orbit) evaluated with the Brown Hamiltonian. The blue and gray lines are evaluations from the N-body integration using the Brown and secular Hamiltonian, respectively.} The salmon solid lines are the boundaries of $E_c^\pm$ from Eq. \ref{Ec_pm}, averaged over the outer orbital period. {Right panels:} Phase portraits for $e_1 = \omega_1$ {space only with $\mathcal{H}_{\rm sec}$ (top), and $\mathcal{H}_{\rm sec} + \mathcal{H}_{\rm B}$ (bottom)}. The grey dots are from the orbital evolution, {the black and green dots are from the secular evolution for $\mathcal{H}_{\rm sec}$ and $\mathcal{H}_{\rm sec} + \mathcal{H}_{\rm B}$, respectively}. The separatrix is the red line, and the 'mean energy' curve from the integrations is the navy line. The initial condition from NASA Horizons data is the purple dot.}
    \label{fig4}
\end{figure*}

\begin{figure*}
    \centering
    \includegraphics[width=1\textwidth]{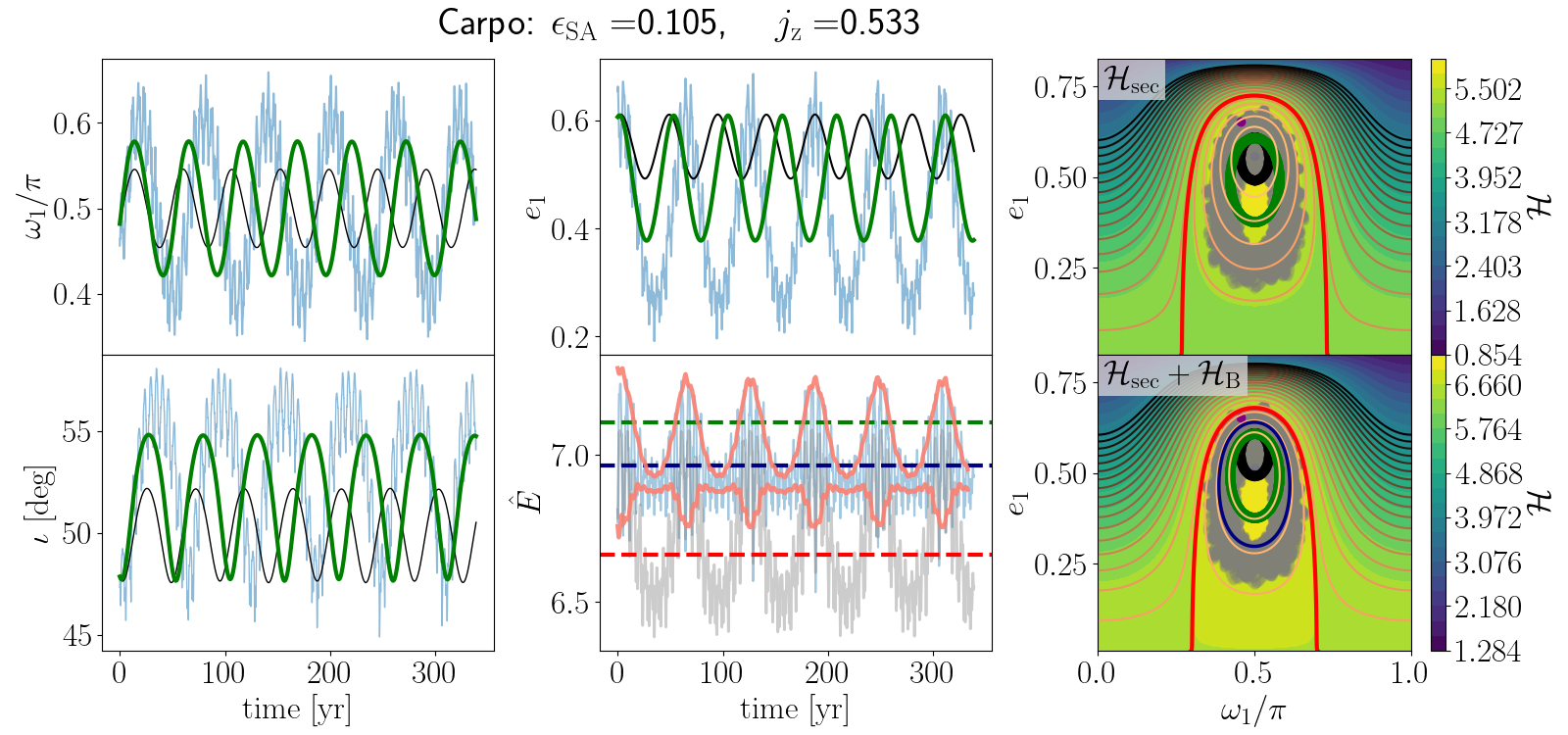}
    \includegraphics[width=1\textwidth]{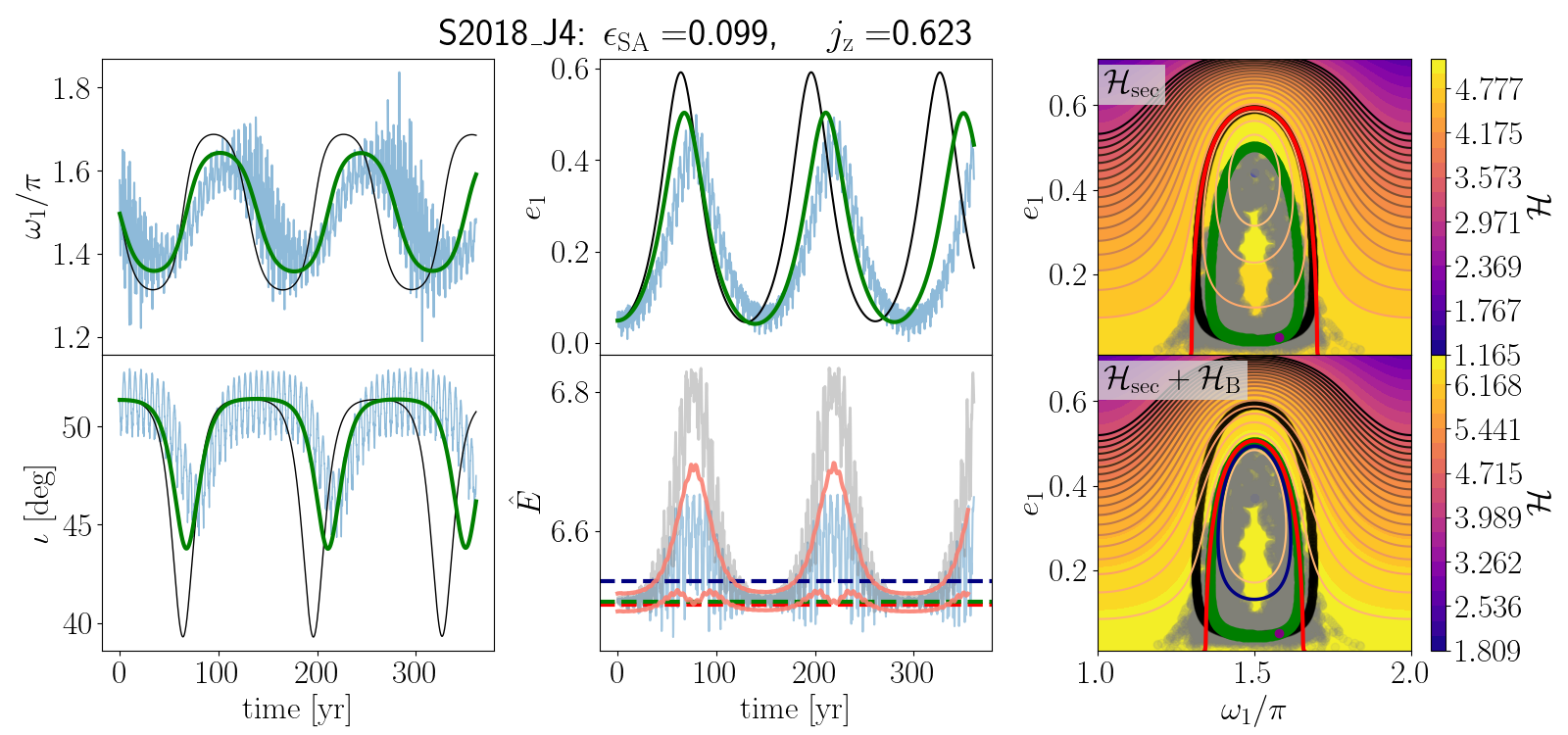}

     \caption{Orbital evolution for Carpo and $S2018\_{\rm J}4$. {The panels are the same as in Fig. \ref{fig4}, but without the evolution of $j_z$ and $\Omega_1$.} }
    \label{fig5}
\end{figure*}

\begin{figure*}
    \centering

    \includegraphics[width=0.93\textwidth]{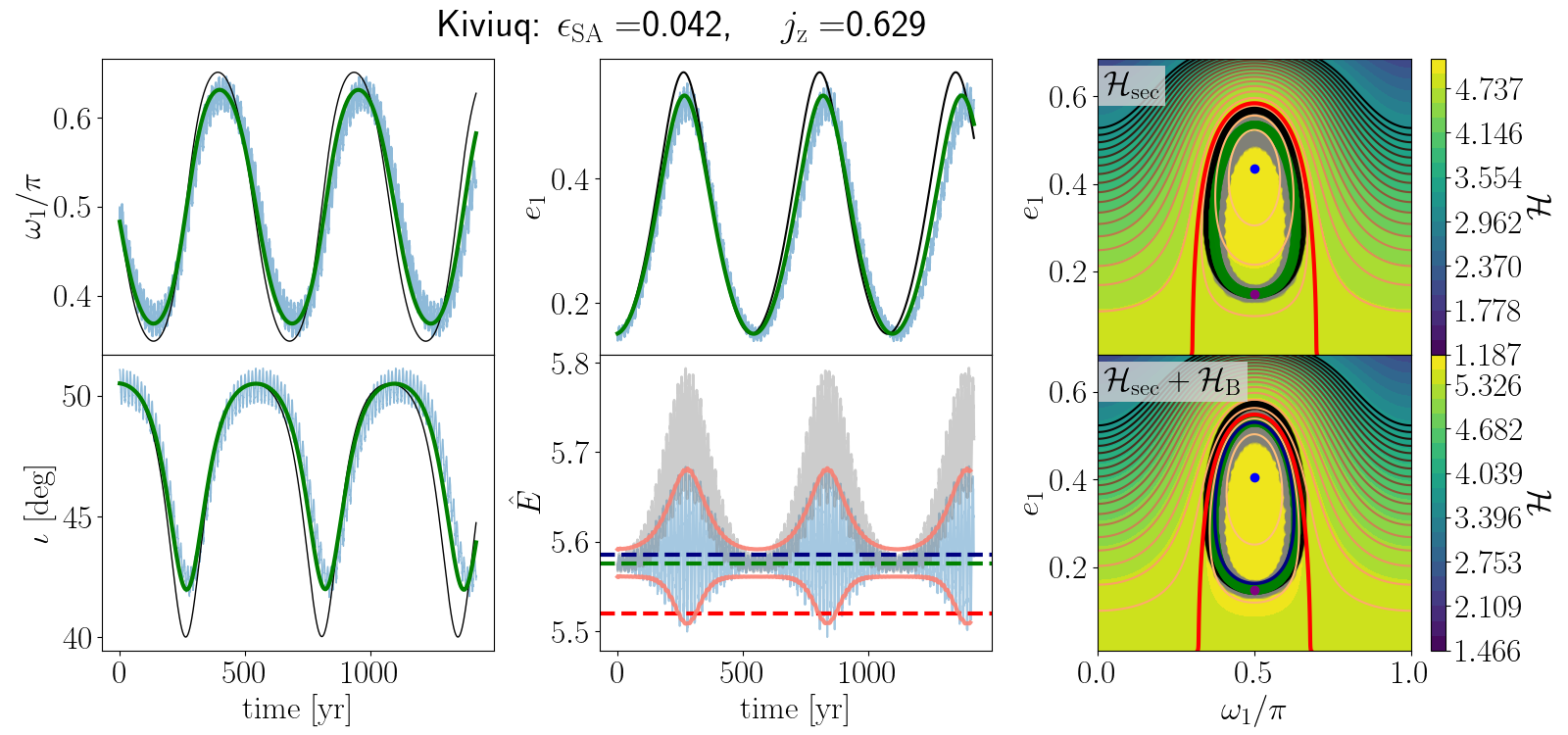}
    \includegraphics[width=0.93\textwidth]{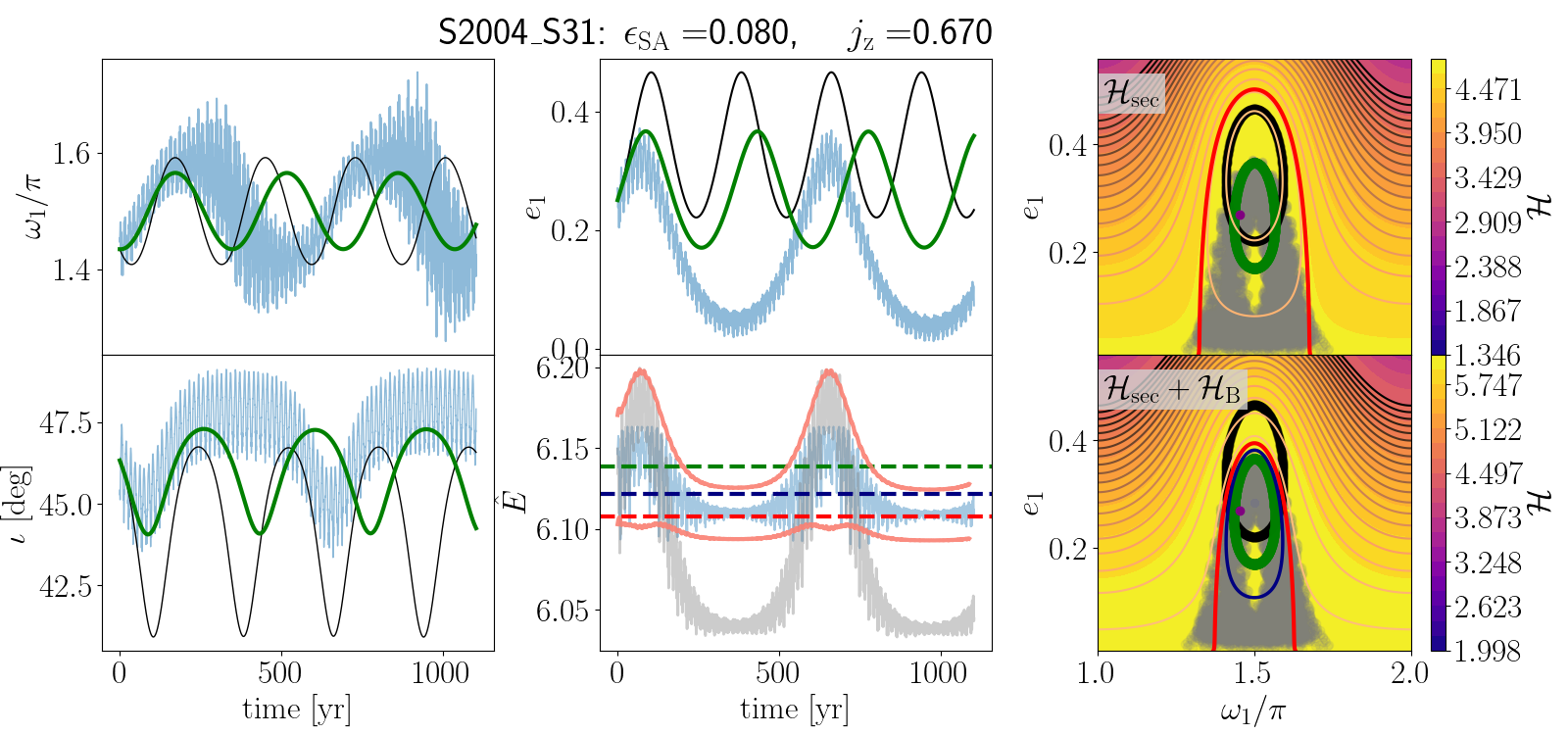}
    \includegraphics[width=0.93\textwidth]{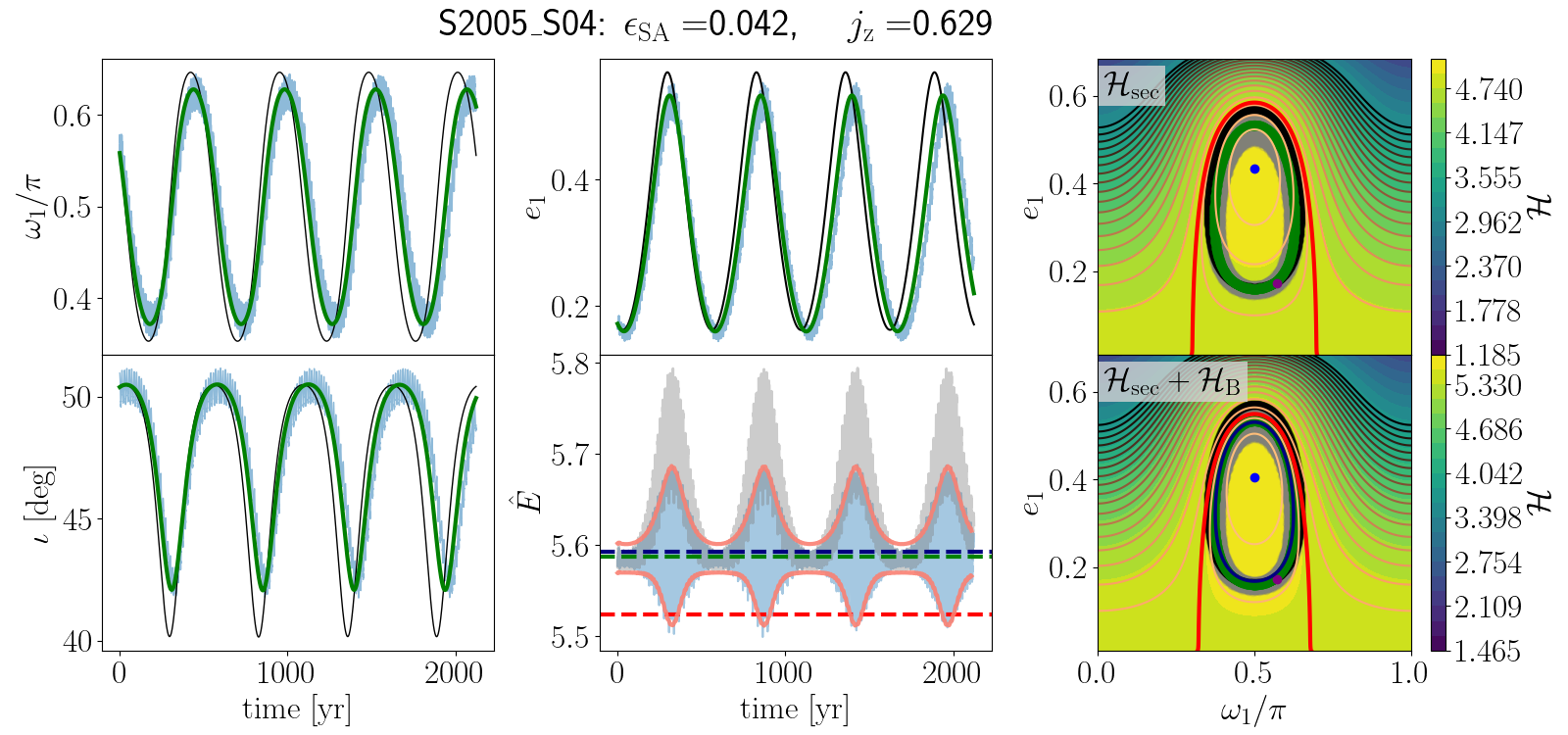}
    
     \caption{Same as Fig. \ref{fig5} but for Saturns' satellties.}
    \label{fig6}
\end{figure*}

Fig. \ref{fig9} shows the dimensionless energy $\mathcal{E}_c$ (Eq. \ref{curlyE}) for each of the suspected satellites together with the boundaries of the fluctuating energy given by $\mathcal{E}_c^\pm$. The condition for being potentially in libration is $\mathcal{E}_c\in[0,1]$ where the extreme values are attained only if the initial energy is exactly at the separatrix (for $0$ - dashed red line) or at the fixed point (for $1$ - dashed green line). 6 out of the 13 librating satellites have always positive energy. Most of the circulating orbits have a large extent of $\mathcal{E}_c$ and can spend a significant amount of time below $0$. 

The boundary of $\mathcal{E}_c^+$ can extend beyond unity because the fluctuation in $j_z$ shifts the Hamiltonian to a different system, and the new $E_c^+$ can exceed the original $E_{\rm fix}$. Of course, $E_c^+$ cannot exceed $E_{\rm fix}^+$, the Energy at the fixed point evaluated for at $j_z + \Delta j_z/2$.

Fig. \ref{fig3} shows the same data in $\mathcal{E}_c^-$ - $\Delta \mathcal{E}$ space where 
\begin{equation}
    \Delta \mathcal{E} = \left| \frac{\mathcal{E}_c }{ \mathcal{E}_c^- } \right| = \left| \frac{E_c - E_s}{E_c^- - E_s}  \right| 
\end{equation}
is the relative extent of the fluctuation. We discard satellites with $\mathcal{E}_c^->0$ since they are always librating. The former measures how deep the fluctuating energy can be below the separatrix while the latter measures the relative extent of the fluctuating energy compared to the initial value. We see that the areas of $\mathcal{E}_c^- < -0.1$ or $\Delta \mathcal{E} < 2$ (dashed red lines) occupied only by librating orbits, while the remaining area, {(coloured in light green)} has all of the circulating orbits and only two librating orbits, which are the retrograde satellites Neso and Euporie that exhibit highly perturbed orbits\footnote{The prograde satellite Carpo could sometimes be in this area, but is rarely realised in practice.}. We will examine individual orbits below. {Small values of $\Delta \mathcal{E}$ indicate that the fluctuation is much larger than $E_c - E_s$, thus it's harder to maintain a librating orbit, while $-\mathcal{E}_C^{-}$ measures the normalised distance from the separatrix, so larger values translate to stronger fluctuations.}

{Of course, direct N-body integrations will provide the true state of the satellite. The advantage of our method is two-fold: It sheds light on the physical nature of the librating/circulation solution and quantifies where the energy condition is accurate and where it is not. Moreover, when the initial conditions are not exactly known and in population studies where exploration of very large initial conditions is required, N-body integrations are less efficient than analytic estimates, which require little computational power.}

{Finally, we stress that although the value of $\epsilon_{\rm SA}$ specifies the strength of the fluctuations (which provides $\mathcal{E}_c^{\pm}$ via $\Delta j_z \propto \epsilon_{\rm SA}$), the question of correctly predicting whether a satellite will librate can not be determined from $\epsilon_{\rm SA}$ alone, and required information on the relative distance from the separatrix (which provides $\mathcal{E}_c$). Satellites with lower $\epsilon_{\rm SA}$ are described well by the Brown Hamiltonian, but as we shall see, they could be misclassified (e.g. $S2019\_S01$, while satellites with larger $\epsilon_{\rm SA}$ are correctly classified (e.g. $S2020\_S05$).}

{To summarise}, we propose the following way to classify librating orbits:

\begin{enumerate}[label*=\arabic*.]

\item From the initial conditions, calculate the energy of the total (secular and Brown) Hamiltonian in Eq. \ref{E_c} and the dimensionless energy $\mathcal{E}_c$ in Eq. \ref{curlyE}. If $\mathcal{E}_c<0$ the orbit is not librating, otherwise:

\item Calculate the fluctuating terms $E_c^{\pm}$ in Eq. \ref{Ec_pm} and $\mathcal{E}_c^{\pm}$ in Eq. \ref{curlyEd}. If the minimum of $\mathcal{E}_c^\pm$>0 is positive, the orbit is librating, otherwise:

\item Calculate the relative energy $\Delta \mathcal{E}=\mathcal{E}_c/(-\mathcal{E}_c^-)$. If $\mathcal{E}_c^- < -0.1$ or $\Delta \mathcal{E} < 2$, the orbit is librating. Otherwise, it is most likely circulating, but direct N body integration is required to confirm.

\end{enumerate}

Generally, prograde satellites tend to be librating is allowed and follow the libration-circulation limits. Retrograde satellites, even though more numerous, tend to have highly perturbed orbits and even their librating orbits are unusual. We examine several individual satellites below.

\begin{figure*}
    \centering

     \includegraphics[width=0.93\textwidth]{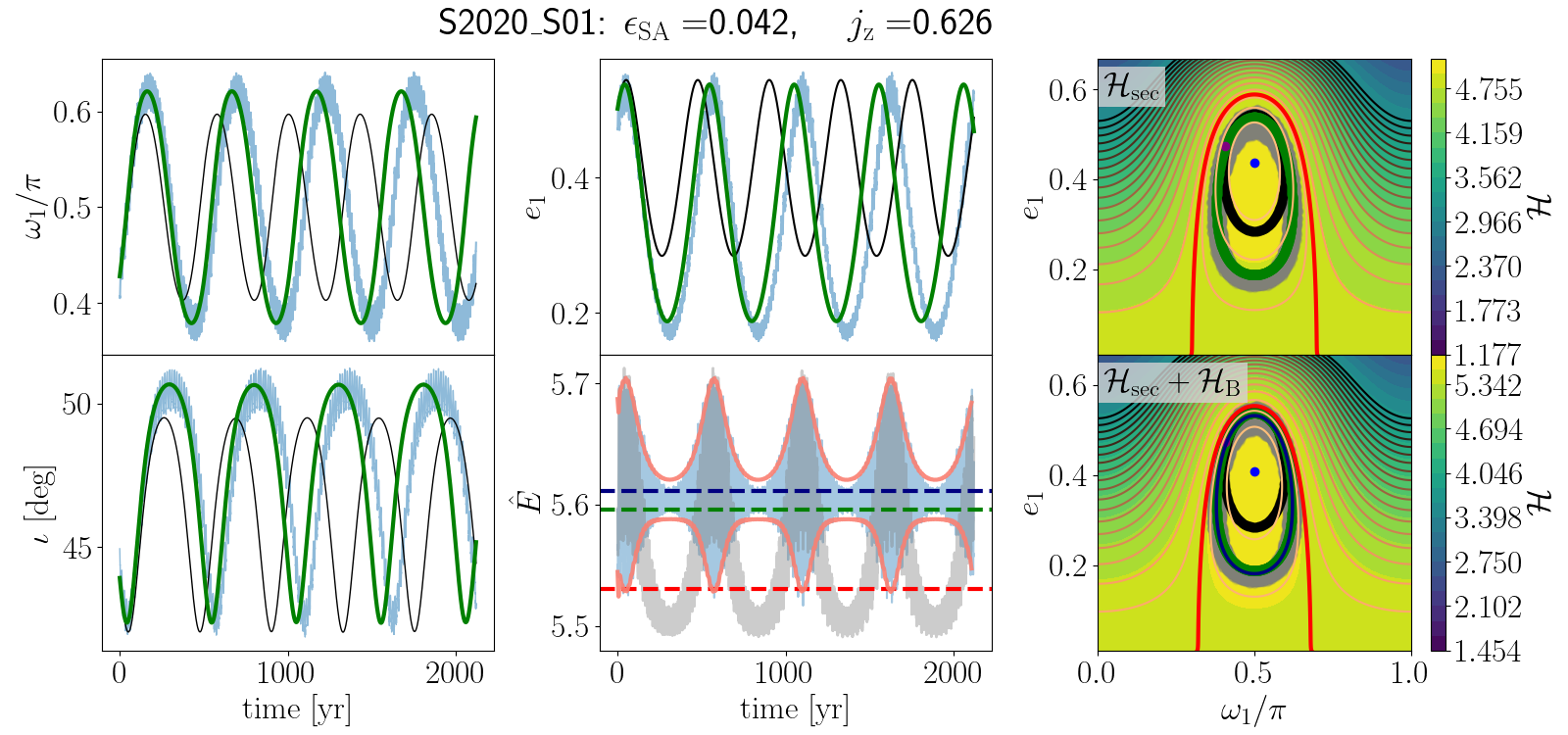}
    \includegraphics[width=0.93\textwidth]{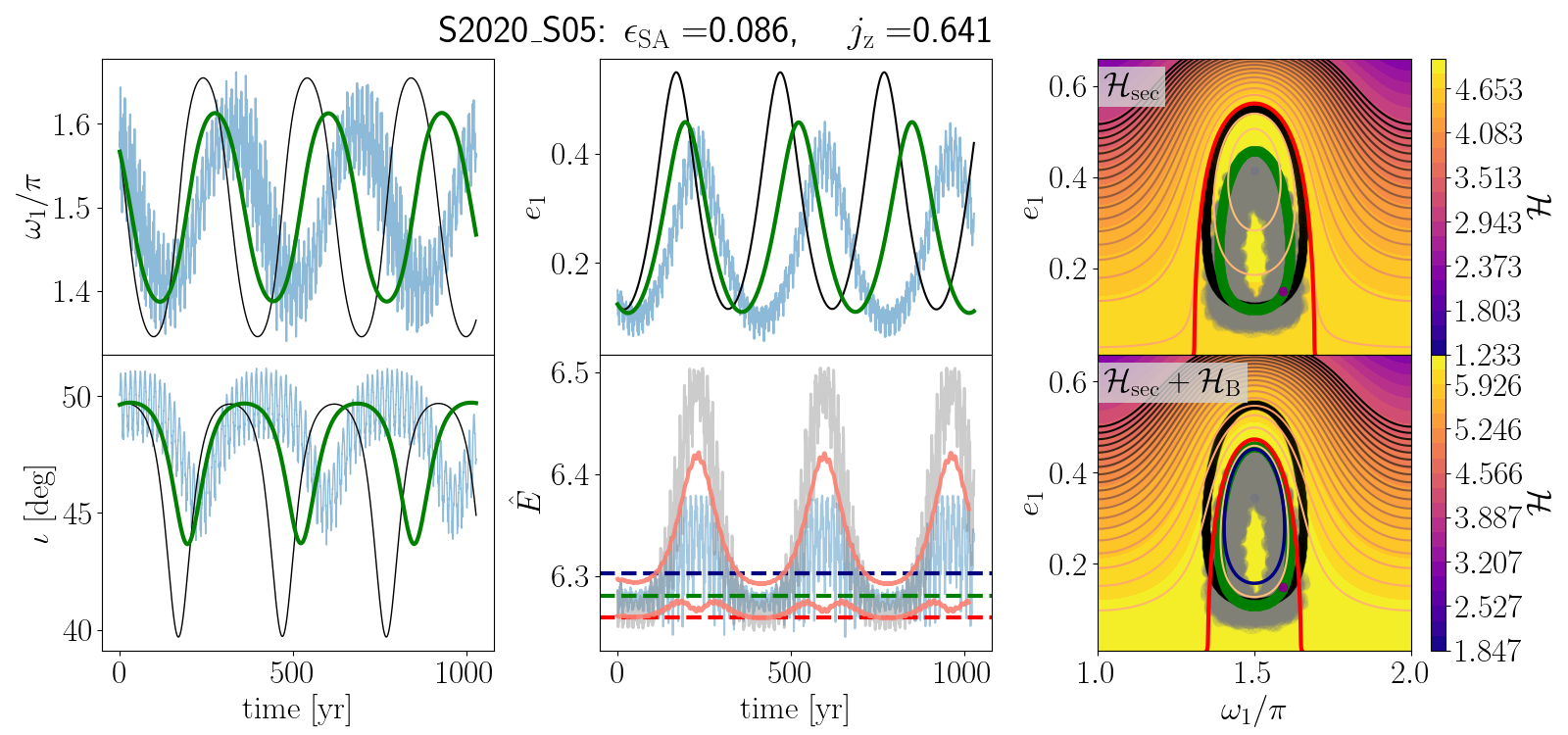}
     \caption{Same as Fig. \ref{fig5} but for different Saturns' satellties.}
    \label{fig7}
\end{figure*}

\begin{figure*}
    \centering

     \includegraphics[width=0.93\textwidth]{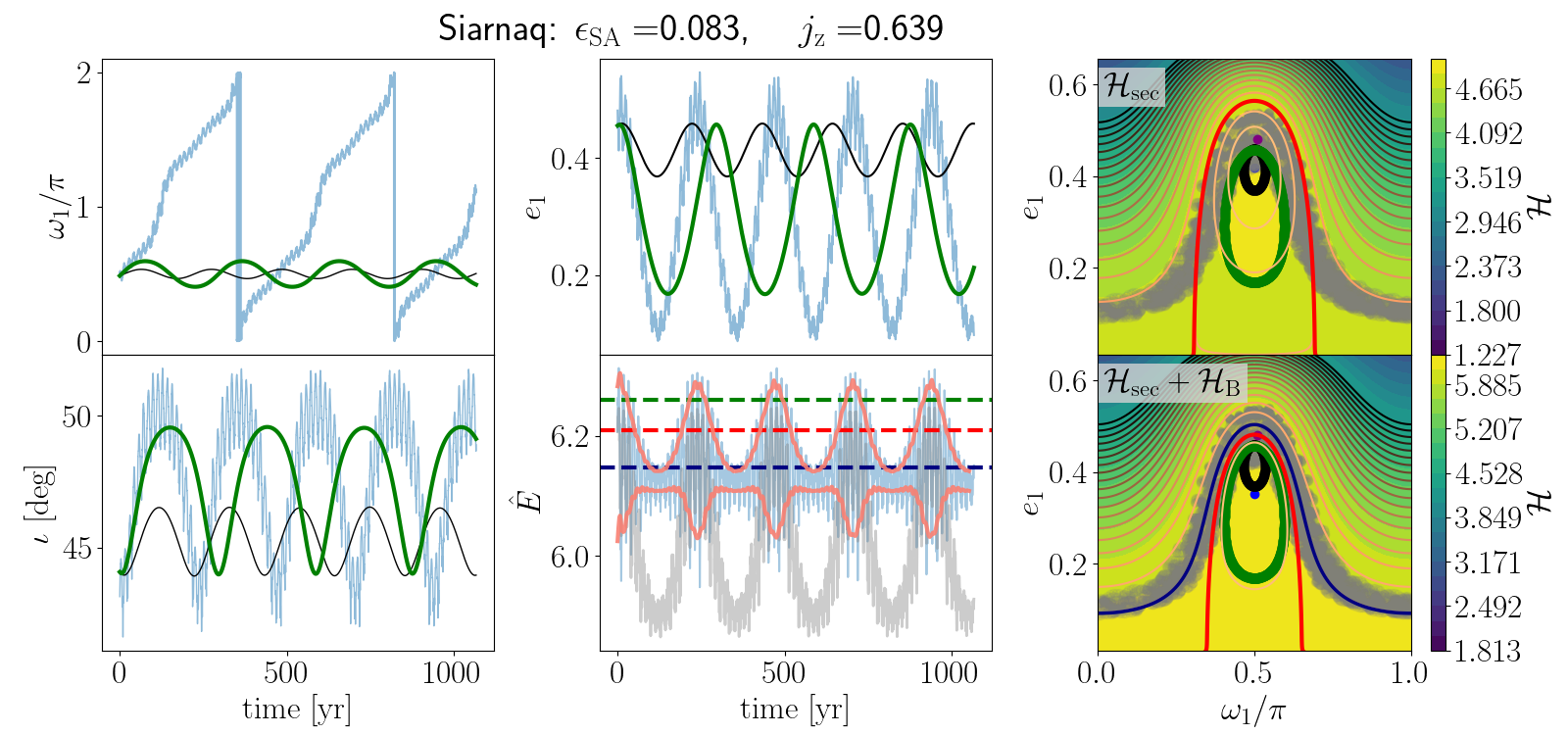}
    \includegraphics[width=0.93\textwidth]{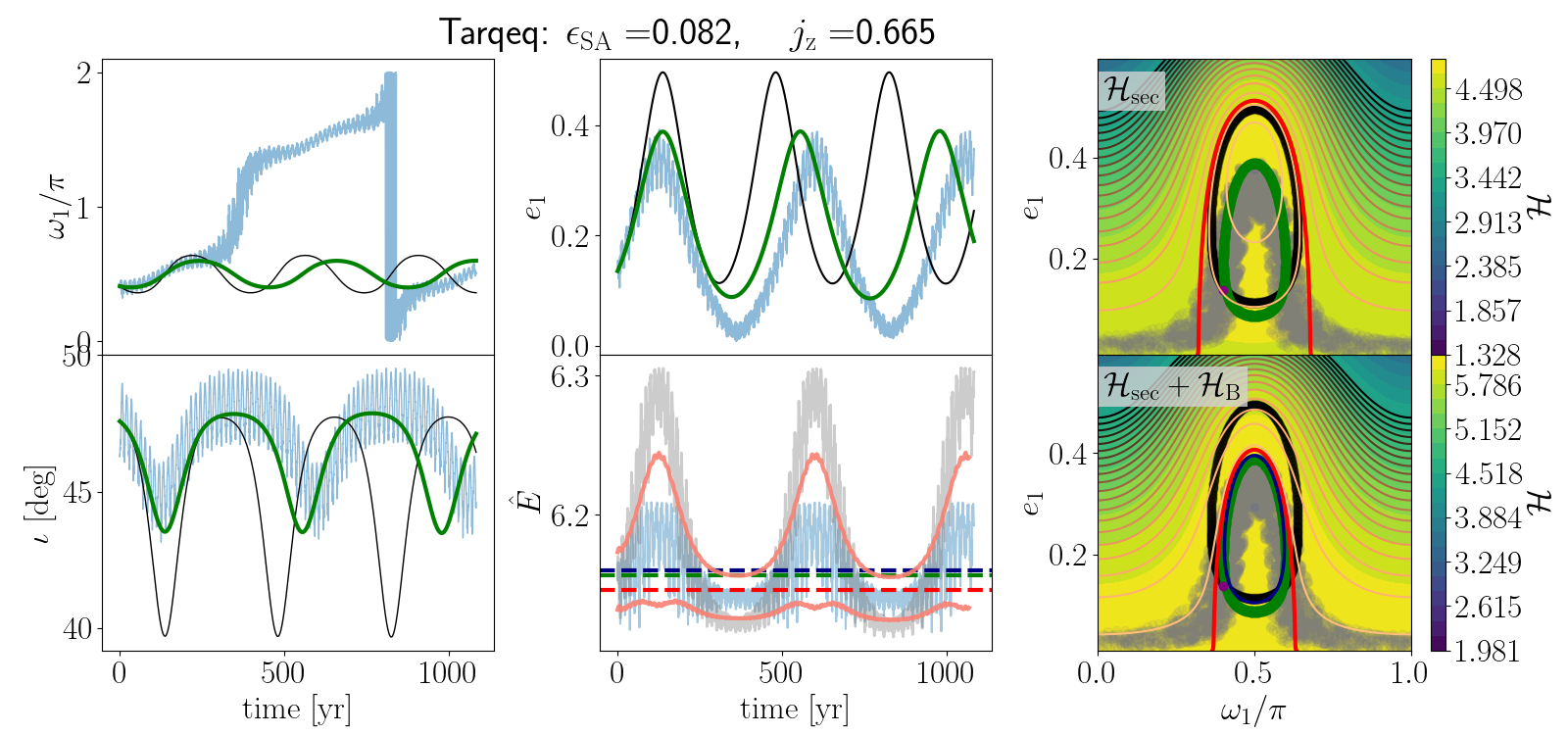}
    \includegraphics[width=0.93\textwidth]{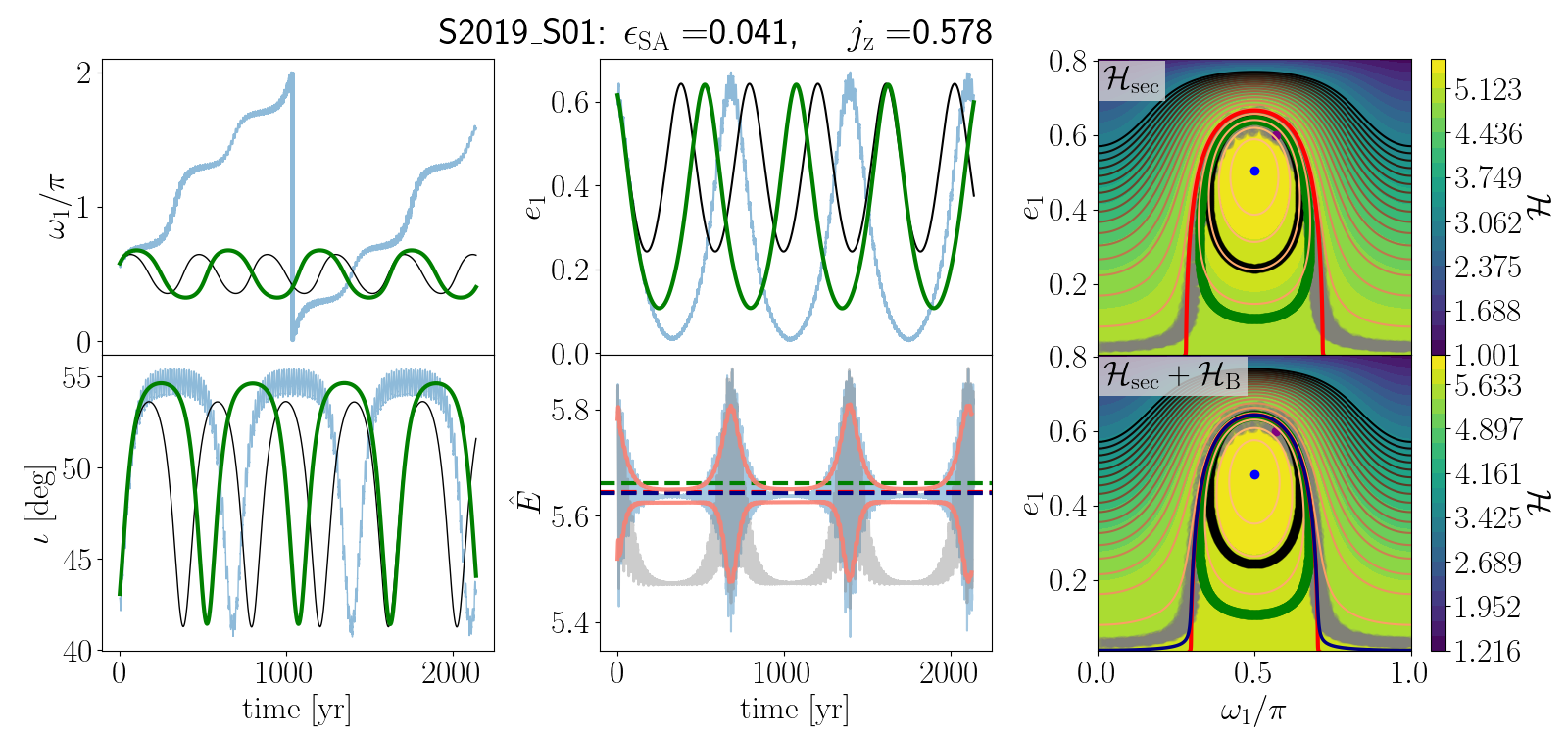}
     \caption{Same as Fig. \ref{fig5} but for circulating Saturns' satellties.}
    \label{fig8}
\end{figure*}

\begin{figure*}
    \centering

    \includegraphics[width=0.93\textwidth]{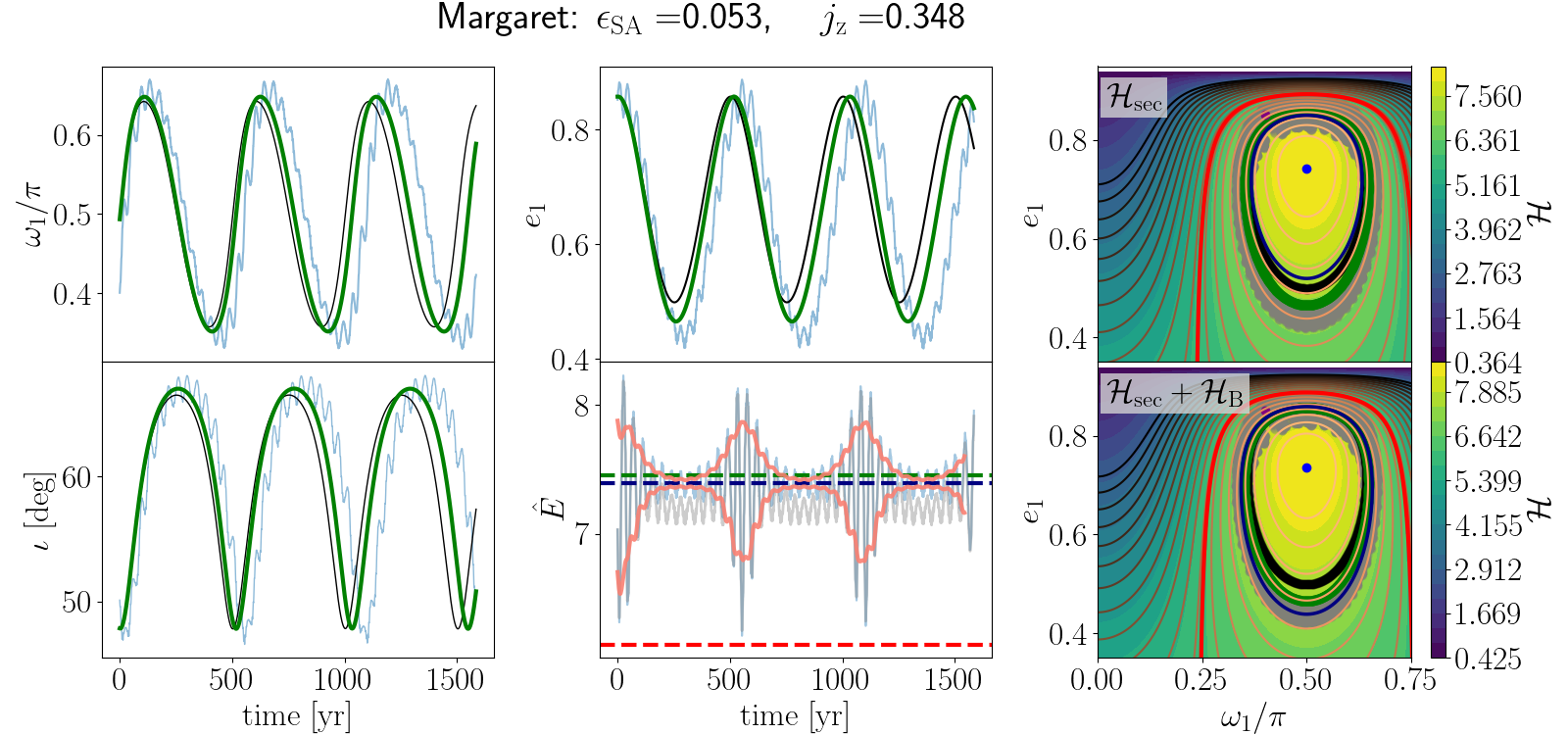}    \includegraphics[width=0.93\textwidth]{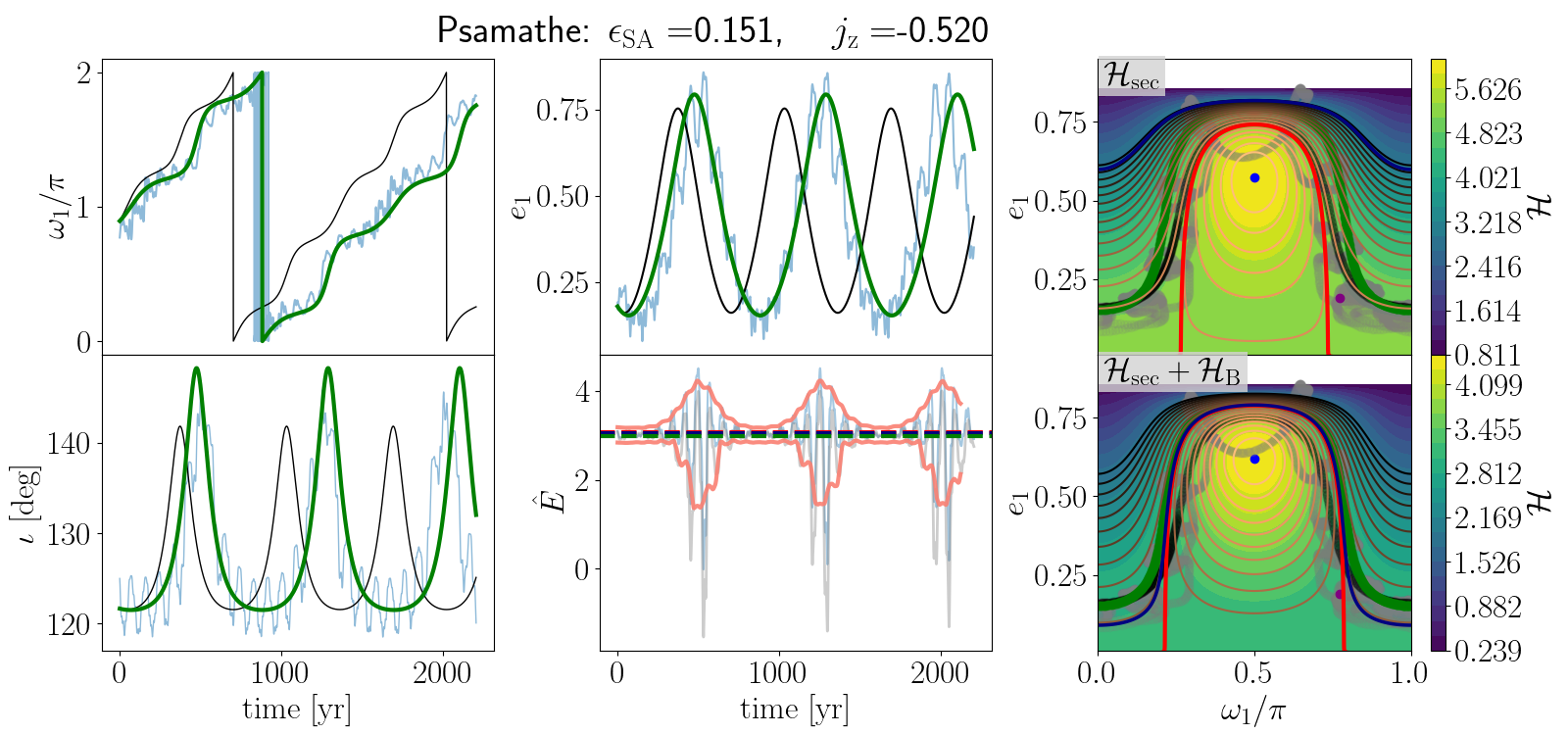}

     \caption{Same as Fig. \ref{fig5} but for Uranus' satellite Margaret and Neptune's circulating satellite Psamathe.}
    \label{fig9}
\end{figure*}

\begin{figure*}
    \centering

    \includegraphics[width=0.93\textwidth]{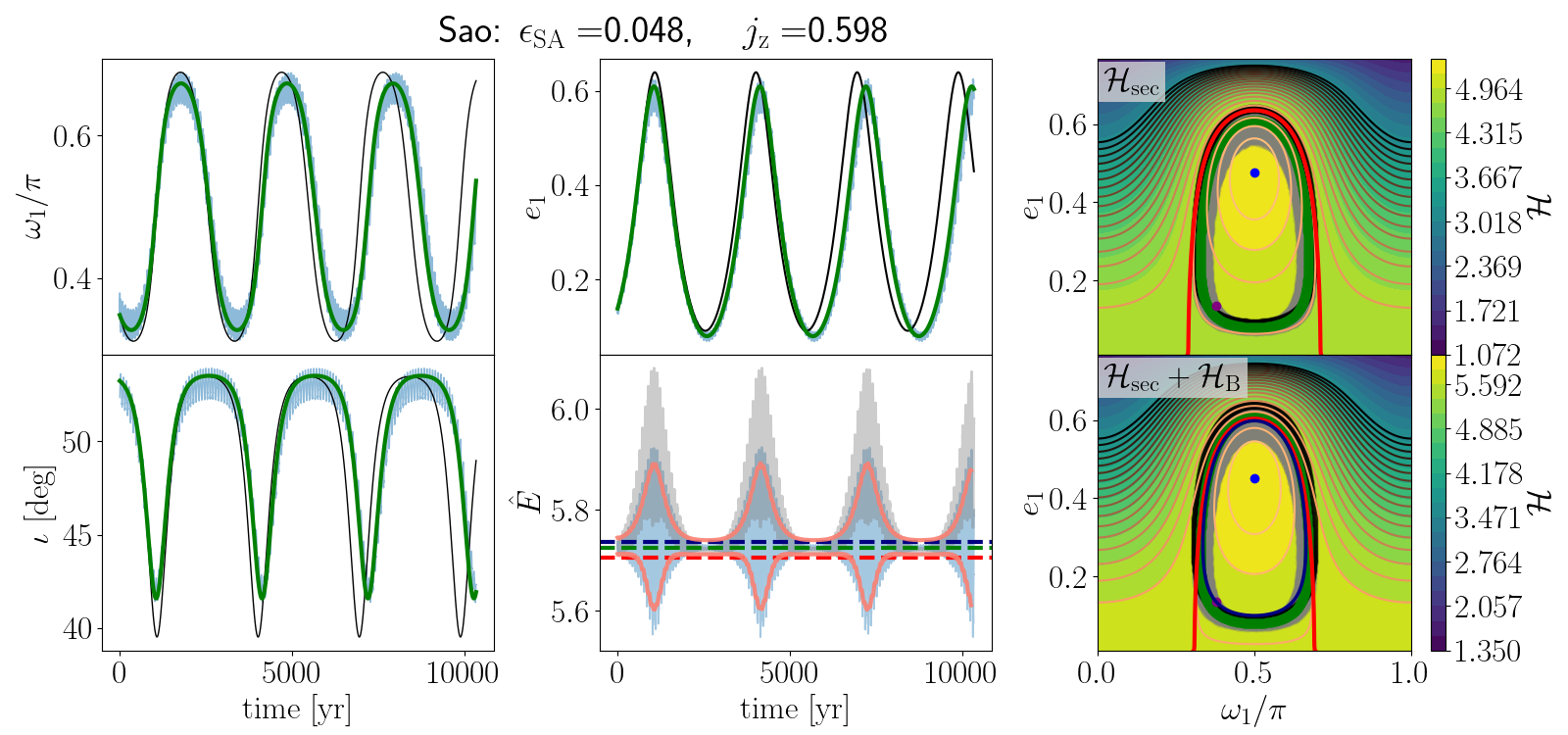}
    \includegraphics[width=0.93\textwidth]{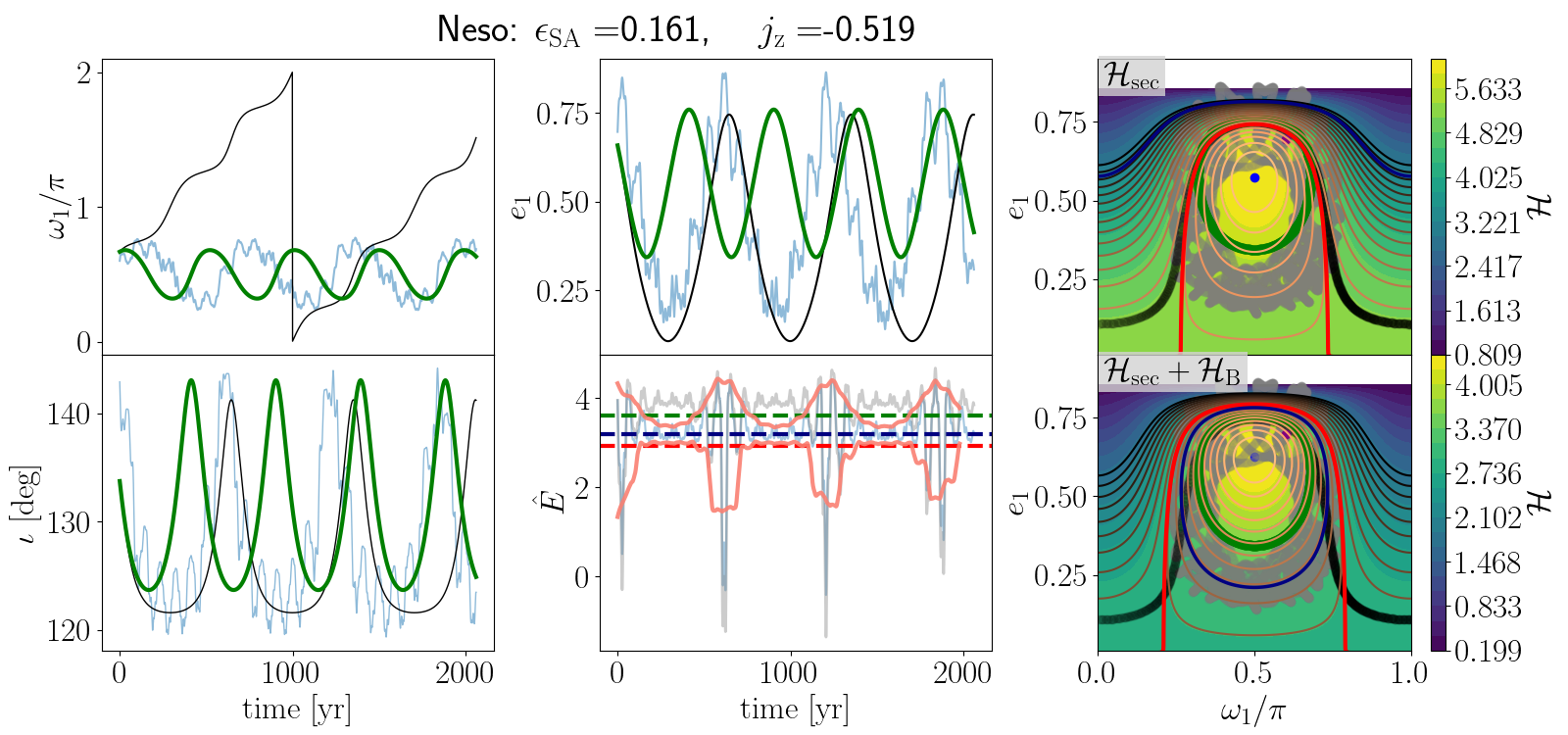}
    \includegraphics[width=0.93\textwidth]{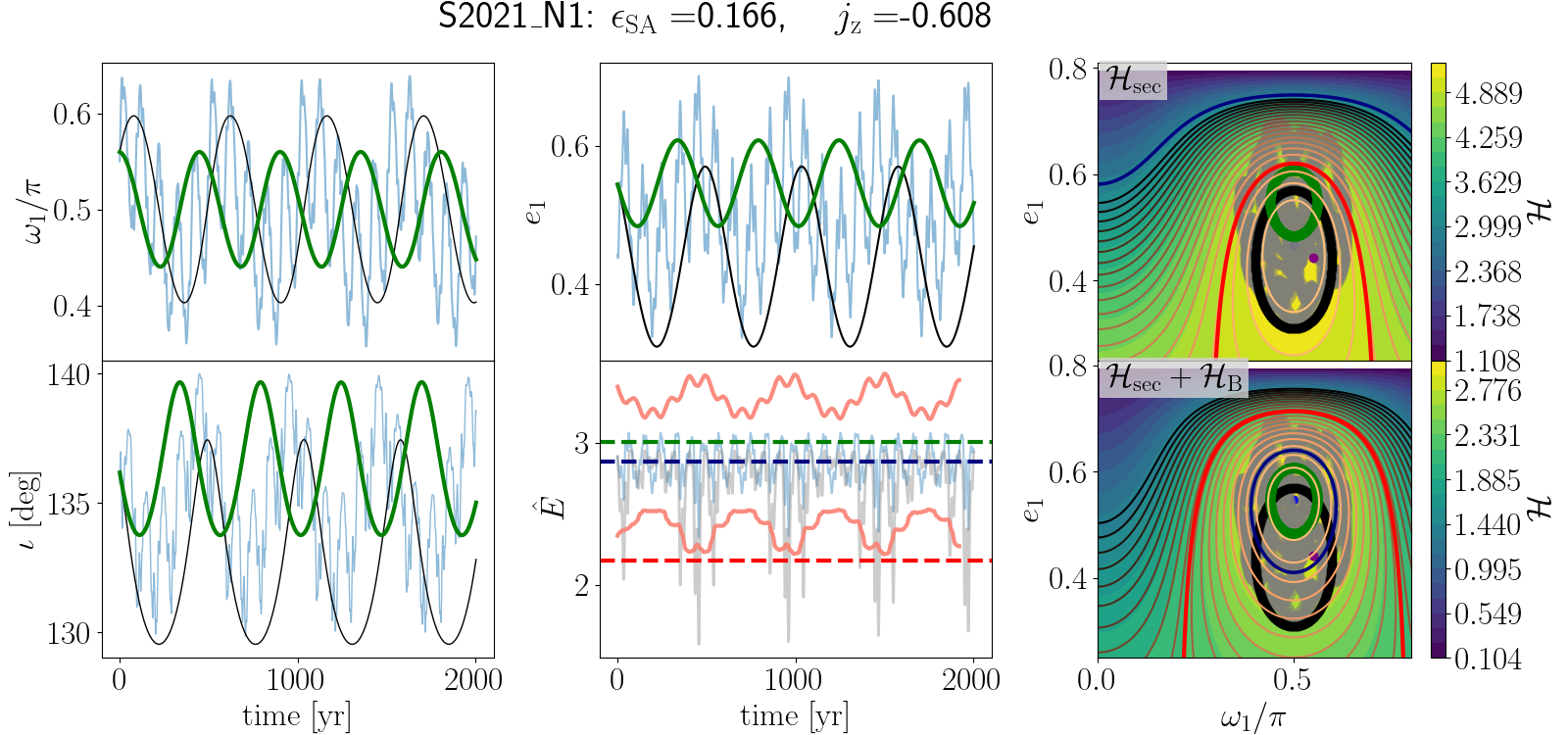}
    
     \caption{Same as Fig. \ref{fig5} but for different Neptune's satellites.}
    \label{fig10}
\end{figure*}

\section{Orbits of individual satellites}\label{s5}
In this section, we closely examine the orbits of some IS. In all the plots the orbital elements are plotted in the invariable plane (i.e. the orbital plane aligns with the Sun-planet orbit). In order to transform to the invariable plane, we first obtain the orbital elements of the planet $\omega_p, i_p, \Omega_p$. Some satellites are close to the separatrix and their fate could be sensitive to the exact initial conditions. {We've tested several satellite orbits with additional giant planets and found no significant difference for the presented timescales.  Long term integration of the full Solar system planets will lead to secular resonances between the planet and satellites' longitude of pericentre precession (e.g., \citealp{st93}), so our numerical integration could deviate for much longer timescales.} Although our simplified three-body integrations agree with past detailed integrations of the Solar system, the ephemeris accuracy decreases over time \citep{em22}. Further observations are required for constant updates of the ephemeris.

 After loading the planet and the satellites' state vectors, we then use \texttt{Rebound}'s built-in rotation framework \href{https://rebound.readthedocs.io/en/latest/ipython_examples/Rotations/}{\texttt{rebound.rotation.orbit()}} which uses quaternions. The Euler angles of the rotation matrix are directly specified to \texttt{rotation.orbit()}. Using the inverse rotation $\boldsymbol{P}_1^{-1}(\omega_p)\  \boldsymbol{P}_2^{-1}(i_p)\ \boldsymbol{P}_3^{-1}(\Omega_p)$ aligns the plane of the planetary orbit with the $XY$ reference plane (see e.g. sec. 2.8 of \citealp{md99}). The inverse rotation is also applied on the state vector of the satellite before the integration.

{In addition to the N-body integrations with \texttt{REBOUND}, we also run two instances of the secular code \texttt{SecuLab} \citep{gp22}, which solves the secular equations of motion and includes the extra terms from \cite{luo16}, as the effective Brown Hamiltonian, for an arbitrary reference frame, as well as the octupole terms. External effects such as tidal dissipation and galactic tides are turned off. We compare direct N-body to the secular terms alone and to the secular and Brown terms. For the initial conditions, we use the N-body integration to generate averaged values of the orbital elements after one inner orbit.}

In generating Figs. \ref{fig4} -- \ref{fig10} the energy in the bottom middle panels is calculated at each timestep $E_c(t)$. The fluctuating amplitudes $E_c^\pm(t)$ (salmon lines) are averaged over the outer orbital period using \texttt{Python}'s \href{https://bottleneck.readthedocs.io/en/latest/bottleneck.move.html}{\texttt{bottleneck.move}} package. The phase portraits (bottom right panels) are similar to the phase portrait in \citetalias{gri24I}.

 \subsection{Jovian Satellites}
 \cite{brozivic17jup} studied the orbits of IS and found that Euporie is librating. We confirm this result in Fig. \ref{fig4} where the orbital elements are similar to their Fig. 6, although we only have three body integrations and not the full model of the Solar System\footnote{Also our orbital elements are in the planet's invariable plane while other studies usually present the orbital elements in the ecliptic reference frame, so small differences are expected.}. Euporie almost does not have any noticeable ZLK cycles, but remains to be librating. The eccentricity remains relatively low and the fluctuations in the energy are overpredicted (which causes the miss-classification of Euporie. A satellite with higher eccentricity at the fixed point would have escaped to circulation, which makes its orbit unusual. The modulation in the energy suggests that other neglected terms may be important and/or the current expansion is inaccurate.
 
 {Fig. \ref{fig5} shows the orbital evolution of Carpo and $S2018\_J4$.} Carpo seems to librate as expected, and the fluctuation in the energy follows the prediction, {however the Brown Hamiltonian model is still not accurate enough.} $\rm S2018\_J4$ was discovered only recently and is not included in the \cite{brozivic17jup} analysis.  Recently it was reported that $\rm S2018\_J4$ has a "similar orbit" to Carpo \citep{sheppard23}, but its dynamical evolution was not explicitly investigated. We suggest that $\rm S2018\_J4$ is also librating on an orbit very close to the boundary. $\omega_1$ seems to have a relatively large amplitude and the orbit spends a significant amount of time beyond the separatrix. Although both Carpo and $\rm S2018\_J4$ librate, their orbits are significantly differ: $\rm S2018\_J4$  librates much slower, its argument of pericenter is librating around $3\pi/2$ and has a larger excess, and the minimal eccentricity attains much lower values.
 
 {To conclude, we see from Jovian satellites that the separatrix boundary in the Brown Hamiltonian is tighter for prograde orbits and correctly displays the actual phase portrait boundary. Even if the Brown Hamiltonian model is not accurate, it is still a much better approximation than the secular Hamiltonian. This is also striking for the retrograde satellite Euporie, where the secular Hamiltonian does not have a fixed point or librating orbits at all, while in the Brown Hamiltonian the orbit is at least contained to the separatrix.}

\subsection{Saturnian satellties}
Saturn has the largest number of IS. \cite{jacobson22} found that three of them are in the ZLK librating mode. We confirm that Kiviuq is in librating mode (top panel of Fig. \ref{fig6}), and its evolution is similar to Fig. 2 in \cite{jacobson22}. The same is true for Ijiraq (not shown) and $S2004\_S31$, where its orbital elements evolve similarly to Fig. 4 in \cite{jacobson22}. Its orbit is qualitatively similar to $S2018\_J4$ with a large amplitude variation of $\omega_1$ and extending well beyond the separatrix. {The correspondence is worse, mainly because the initial energy estimated from the Brown Hamiltonian $\langle E_{\rm B} \rangle$ is further from the separatrix than in reality.}

\cite{sheppard23} suspected that $\rm S2005\_S04$ could be part of the Inuit group, but it has a significantly higher inclination. Indeed, the instantaneous inclination to the ecliptic is $52.6$ degrees, $\sim 3$ degrees larger than the rest of the Inuit group. However, when we rotate to the invariable plane, the orbit of $\rm S2005\_S04$ is very similar to Ijiraq, and both satellites are in ZLK libration mode. Because the longitude of ascending nodes $\Delta \Omega_1$ of the two satellites is quite different, it translated to a noticeable difference in the ecliptic reference frame. 

Besides the known librating satellites and the suggestion that  $S2005\_S04$ is also an Inuit-group librating satellite, we also propose that $S2020\_S01$ and $S2020\_S04$ are librating (Fig. \ref{fig7}). $S2020\_S01$ is part of the Inuit group and shared similar orbital elements with Kiviuq and Ijiraq \citep{sheppard23}, but it was not explicitly mentioned that it is librating.  

{In addition, Saturn has a set of satellites that are very close to the boundary, but are circulating. Two of them ( $\rm S2019\_S01$ and Tarqeq) appear in Fig. \ref{fig1} as suspected, while Siarnaq's current initial condition is just outside the separatrix and is not classified as suspected. We see in Fig. \ref{fig8} that although the Brown Hamiltonian secular code (green lines) captures the long-term evolution better, the classification is inaccurate since both secular codes predict a librating orbit. Nevertheless, the boundary of the separatrix from the Brown Hamiltonian fits much better to the orbital evolution.}

\subsection{Uranian Satellites}
The only prograde irregular satellite of Uranus is Margaret, which is also the only one librating \citep{brozovic09, brozovic22}. The top panel of Fig. \ref{fig9} depicts Margaret's orbital evolution and is consistent with Fig. 24 in \cite{brozovic09} and Fig. 5 in \cite{brozovic22}. Margaret by far has the lowest $j_z$ value and can reach high eccentricities up to $\sim 0.9$. Margaret has the strongest resonance of all the prograde satellites in the sense that it is the furthest from the separatrix in dimensionless $\mathcal{E}_c$ energy units.  

\citetalias{gri24I} found a potential hypothetical family of highly inclined, highly eccentric, librating orbits that can extend far beyond the standard stability limit. This was also speculated by \cite{car02} on similar grounds. Margaret's $\epsilon_{\rm SA}$ is relatively small for such a population, but it can still be a representative member of such a population due to its highly eccentric and stable orbit. 

\subsection{Neptunian satellites}
Neptune has two known librating satellites, Sao and Neso \citep{brozovic11}. While Sao is prograde and has a relatively standard orbit, similar to Saturn's Inuit group, Neso is located further out and has a retrograde highly perturbed orbit. Our integrations are shown in Fig. \ref{fig10} and are consistent with Fig. 5 of \cite{brozovic22}. {Note that the ZLK Hamiltonian misclassified Neso's orbit and the Brown Hamiltonian is required to correctly recover its librating orbit, even though neither of the secular models is accuracy enough.}

The outermost satellite is $\rm S2021\_N1$, discovered only recently\footnote{The official announcement was made on 23 February 2024: \href{https://minorplanetcenter.net/mpec/K24/K24DB2.html}{ https://minorplanetcenter.net/mpec/K24/K24DB2.html}.}. We show that it is also in ZLK libration\footnote{The Wikipedia page for $\rm S2021\_N1$ states that the argument of pericentre librates around $90$ deg with a reference to \cite{brozovic22}, but $\rm S2021\_N1$ is not mentioned or included in their analysis.} (bottom panel of Fig. \ref{fig10}, but the orbit slightly differs from Neso. While Neso avoids reaching high energies and its phase space is 'doughnut shaped', $\rm S2021\_N1$ easily covers the phase space's fixed point. On the other hand, its energy fluctuations are much smaller, and it crosses the separatrix only for short periods in the highly eccentric phases. The argument or pericentre also has a smaller libration amplitude compared to Neso, which is attributed to lower $j_z$ values. {Using the ZLK Hamiltonian completely misses $\rm S2021\_N1$'s phase portrait, and the additional of the Brown Hamiltonian makes it contained within the separatrix.}

The bottom panel of Fig. \ref{fig9} shows the orbital evolution of the circulating satellite Psamathe, which has a highly perturbed and unusual trajectory in phase space. {Neither of the secular models can describe Psamathe's orbital evolution.}

\section{Summary and conclusions}\label{s6}
We studied analytically and numerically the long-term evolution of triple systems of mild hierarchy, where the period ratio between the inner and outer orbits is not too small. Our conclusions are summarised below:

\begin{itemize}

\item We find that all of the librating satellites except Euporie are within the zone allowed for libration (Fig. \ref{fig1}). Most of the librating satellites are prograde and they can be very close to the librating-circulating boundary (Eq. \ref{eps_fit}).

\item We formulate a criterion that predicts whether a satellite's argument of pericentre will be librating based on its initial conditions (Fig. \ref{fig2}). The criterion utilises the effective energy conservation given by the secular and Brown Hamiltonians (Eq. \ref{E_c}) and the magnitude of the osculating energy (Eq. \ref{Ec_pm}). 

\item While the prograde satellites mostly follow the simplified criterion (sec. \ref{s41}), the highly perturbed IS require a more refined criterion (sec. \ref{s42}). The exact process is described at the end of sec. \ref{s4}. Using this criterion we were able to confirm that all previously known IS are librating. In addition, we find that $\rm S2018\_J4$,  $\rm S2005\_S04$, $\rm S2020\_S01$, $\rm S2020\_S04$, and $\rm S2021\_N1$ are also librating. Future observations and missions would increase our sample and test our predictions.

\item Although retrograde orbits are more numerous and have more available parameter space, the vast majority of librating satellites are prograde. The extra energy from the Brown Hamiltonian (and its osculating boundaries) correctly predicts the satellite's motion in the $e_1-\omega_1$ phase space (Figs. \ref{fig4}-\ref{fig10}), except Euporie, who's osculating magnitude is greatly overestimated.

\item {Even if the evolution is inaccurate, the librating satellites are contained within the libration zonea (right panels of Fig. \ref{fig4} - \ref{fig10}), and the circulating satellites are outside the libration zones right panels of Fig. (\ref{fig8}). }

\end{itemize}

{Future work will be able to better model the retrograde satellites where the Brown Hamiltonian is inaccurate. One option is to construct a simplified version of a higher-order theory \citep{lei19} where the first averaging is done after the canonical transformation. Simplified models will not require literal expansions and will converge to any eccentricity. Future data on IS and binary minor planets in the asteroid and Kuiper belt will put our theory under examination. Additional perturbations from other bodies or tidal and rotational forces could potentially change our results. }

\section*{Acknowledgements}
EG thanks Alessandro Trani, Isobel Romero-Shaw, Mor Rozner, Rosemary Mardling and Scott Tremaine for stimulating discussions and comments on the manuscript. EG thanks Marina Brozivi{\'c} and Hanno Rein for advice on how set up the initial conditions for the N-body integrations. EG acknowledges support from  ARC grant FT190100574 (CI: Mandel).

\section*{Data Availability}
The data used in this paper is public and the methods are publicly available in the \href{https://github.com/eugeneg88/fixed_points_brown}{\texttt{GitHub} repository}.
\bibliography{sample631}
\bibliographystyle{mnras}


\bsp	
\label{lastpage}
\end{document}